\documentclass[a4paper,10pt]{article}
\usepackage[position=t,singlelinecheck=off]{subfig}
\usepackage{amssymb,amsmath,graphics,graphicx,float,braket}
\def\beq{\begin{equation}}
\def\eeq{\end{equation}}
\def\bea{\begin{eqnarray}}
\def\eea{\end{eqnarray}}
\def\nn{\nonumber}

\makeatletter
\def\@cite#1#2{${\mbox{#1\if@tempswa , #2\fi}}$}
\makeatother

\begin{document}
\thispagestyle{empty}
\begin{center}
\begin{LARGE}
\textsf{Evolution of a hybrid micro-macro entangled state of the  qubit-oscillator system via the 
generalized rotating wave approximation}
\end{LARGE} \\

\bigskip\bigskip
R. Chakrabarti$^{\dagger}$ and V. Yogesh$^{\ddagger}$
\\
\begin{small}
\bigskip
\textit{
 $^{\dagger}$Chennai Mathematical Institute, H1 SIPCOT IT Park, \\Siruseri, Kelambakkam 603 103, India\\
$^{\ddagger}$ Department of Theoretical Physics, 
University of Madras, \\
Maraimalai Campus, Guindy, 
Chennai 600 025, India \\}
\end{small}
\end{center}

\vfill
\begin{abstract}
  We study the evolution of the hybrid entangled states in a bipartite (ultra) strongly  coupled qubit-oscillator system. 
 Using the generalized rotating wave approximation the reduced density matrices of the qubit and the oscillator are 
 obtained. The reduced density matrix of the oscillator yields the phase space quasi probability distributions 
 such as the diagonal $P$-representation, the Wigner $W$-distribution and the Husimi $Q$-function. In the strong coupling 
 regime the $Q$-function evolves to uniformly separated macroscopically distinct Gaussian peaks representing `kitten' 
 states at certain specified times that depend  on  \textit{multiple} time scales present in the interacting system. For 
 the ultra-strong coupling realm  a large number of interaction-generated modes arise with a complete randomization of their
 phases. A stochastic averaging of the dynamical quantities  sets in  while leading to the decoherence of the system.  
 The delocalization in the phase space of the oscillator is studied by using the Wehrl 
 entropy.  The negativity of the $W$-distribution, while registering its departure from the classical states, allows us 
 to compare the information-theoretic measures such as the Wehrl entropy with the Wigner entropy. Other features of 
 nonclassicality such as the existence of the squeezed states and appearance of negative values of the Mandel parameter
 are realized during the course of evolution of the bipartite system. In the parametric regime 
 studied here these properties do not survive after a time-averaging process.
\end{abstract} 
 
\newpage
\setcounter{page}{1}

 \section{Introduction}
The quantum mechanical model that describes a  two-level system (qubit) interacting with  an electromagnetic field 
mode (oscillator) in a cavity   possesses a rich theoretical structure.  Its dynamical features, particularly  in the regime of 
a weak  qubit-oscillator coupling and also for a small detuning between the qubit and the oscillator frequencies, are described 
by the exactly solvable Jaynes-Cummings model [\cite{JC1963}] that employs the rotating wave approximation. Recently, 
however, the stronger coupling domain, where the rotating wave approximation no longer holds, has been experimentally 
investigated. Various novel realizations such as a nanomechanical resonator capacitively coupled to a Cooper-pair box 
[\cite{ABS2002}], a quantum semiconductor microcavity undergoing excitonic transitions [\cite{ALT2009}], a flux-biased 
superconducting quantum  circuit that uses large nonlinear inductance of a Josephson junction to produce ultra-strong 
coupling with a coplanar waveguide resonator [\cite{ND2010}] fall in this category. The superconducting qubits and 
circuits, in specific, offer much flexibility in selecting the control parameters. Consequently, they are regarded as suitable 
candidates for the quantum simulators [\cite{BAN2011}-\cite{GAN2014}]. Moreover, integrated hybrid quantum circuits involving 
atoms, spins, cavity photons, and superconducting qubits with nanomechanical resonators have triggered much interest [\cite{ZAYN2013}].

\par

To study the strongly interacting qubit-oscillator system where the Hamiltonian incorporates terms that do not 
preserve the total excitation number, the authors of Refs. [\cite{IGMS2005}, \cite{AN2010}]  have advanced an adiabatic 
approximation scheme that holds in the parametric realm where the oscillator frequency dominates the qubit frequency.
The resultant separation of  the time scales facilitates decoupling of the full Hamiltonian into sectors related to each 
scale, and allows approximate evaluation [\cite{IGMS2005}] of  the eigenstates of the system. To augment the parametric 
domain so that it includes the exact resonance as well as the large detuning regime, a generalization of the rotating wave 
approximation has been proposed [\cite{I2007}]. This generalization utilizes the basis states obtained via the 
adiabatic limit, and subsequently, the excitation number conserving argument \textit{\`{a} la} the rotating wave approximation 
is applied [\cite{I2007}] to the Hamiltonian in the new basis. The resultant block diagonalized Hamiltonian 
now furnishes [\cite{I2007}] energy eigenvalues which are approximately valid for the (ultra) high coupling strength and also
for a wide range of the detuning parameter. 

\par

The hybrid entangled states of the bipartite coupled qubit-oscillator system  provide instances of the entanglement of the 
microscopic atomic states and, say, the photonic  Schr\"{o}dinger cat states  that may be regarded as macroscopic for reasonably large 
values of the coherent state amplitude. These states  play crucial roles in many areas. For instance, they facilitate the 
non-destructive measurement [\cite{Haroche2007}] of the photon number in a field stored in a cavity. Bell inequality tests involving 
these qubit-field entangled states have been proposed [\cite{PSSAJ2012}]. The hybrid entanglement involving a discrete and a continuous 
quantum variable may offer advantages such as achieving near-deterministic quantum teleportation [\cite{LJ2013}]. Moreover, the quantum 
bus [\cite{VL2008}] framework utilizes the hybrid entanglement where direct qubit-qubit interactions are eliminated while providing a 
commonly coupled continuous mode that mediates  among qubits.  Realization of the micro-macro entangled states via a controllable 
interaction  of a single-mode microwave cavity field with a superconducting charge qubit, and the subsequent creation of  the 
superposition of  macroscopically distinguishable field modes by virtue of  measuring  the charge states of the qubit have been 
proposed [\cite{LWN2005}]. Much current experimental activity [\cite{A2014}] is  focused towards generation of such hybrid entanglement. 
Recently the optical hybrid entanglement has been observed [\cite{Jeong2014}] by the quantum superposition of non-Gaussian operations on 
distinct modes. Following a measurement based procedure, the hybrid entanglement between two remote nodes residing in the Hilbert spaces 
of different dimensionality has been established [\cite{Morin2014}]. The superconducting circuits have also been used for controllable 
and deterministic generation of complex superposition of states [\cite{LWN2004}-\cite{Hofheinz2009}].

\par

Our objective in the present work is as follows. Within the generalized rotating wave approximation scheme we analytically 
study the evolution of an initial hybrid entangled state in a coupled qubit-oscillator system. Our study includes both  
the strong and the ultra-strong coupling domains. Tracing over the complementary degree of freedom we obtain the time-evolution 
of the reduced density matrices of the qubit and the oscillator, respectively. The qubit reduced density matrix   provides 
the von Neumann entropy of the system that measures the entanglement and the  mixedness of the state. The oscillator reduced 
density matrix, in turn, yields the phase space quasi probability distributions [\cite{Schleich2001}] such as the diagonal 
$P$-representation, the Wigner $W$-distribution, and the Husimi $Q$-function. The  Wehrl entropy [\cite{W1978}] constructed
via the $Q$-function  measures the delocalization  of the oscillator  in the phase space. In the  strong coupling regime 
where a quadratic approximation to the effective interaction described by the Laguerre polynomials holds, a  long-range 
quasi periodic time  dependence is visible for the Wehrl entropy. This is similar to its behavior  in  the Kerr-like 
nonlinear self-interacting photonic models, where it has been observed  [\cite{MTK1990}-\cite{MBWI2001}] that an initial 
coherent state therein evolves, at rational fractions of the time period, to the superposition of certain macroscopic coherent states 
popularly known as `kitten' states. While paralleling  this phenomenon the current qubit-oscillator interacting model 
has, however, an important distinction that stems from the existence  of interaction-generated  multiple time scales. Quantum 
fluctuations at shorter time scale that signifies the energy exchange between the qubit and the oscillator cause, at rational 
fractions of the long-range time period, a doubling of the number of coherent state peaks (`kittens') while the Wehrl entropy 
evolves from a local minimum to a local maximum of the short-range undulations. A frequency modulation of the short time scale oscillation is also observed. 
In the ultra-strong coupling limit the randomization of the  phases of a large number of incommensurate quantum modes enforces 
a statistical equilibrium, and a stabilization of the occupation on the phase space sets in. Another object of study is the Wigner 
entropy [\cite{SKD2012}] that probes the quantumness of the mixed states linked with the negativity of the $W$-distribution. Our comparison 
 between the Wigner entropy and the Wehrl entropy reveals the close connection of their interrelation with the negativity of the 
 $W$-distribution. Quantum fluctuations in the strong coupling domain induce formations of \textit{almost pure} squeezed states. 
 Using the Mandel parameter [\cite{M1979}] the  nonclassicality of the photon statistics is also observed.

\section{The Hamiltonian and the reduced density matrices}
In natural units ($\hbar=1$) the  Hamiltonian [\cite{IGMS2005}-\cite{I2007}] of the coupled qubit-oscillator system reads 
\beq
H = \omega a^{\dag} a +\tfrac{1}{2} \Delta \sigma_x 
+ \lambda \sigma_z (a^{\dag} + a),
\label{qOH}
\eeq
where the field mode of frequency $\omega$ is described by the annihilation and creation operators 
($a, a^{\dagger}| \hat{n} \equiv a^{\dagger} a$), and the two-level atom having a  transition frequency $\Delta$ is expressed 
via the spin variables $(\sigma_{x}, \sigma_{z})$. The qubit-oscillator coupling strength 
equals $\lambda$.  The Fock states 
$\{\hat{n} |n\rangle = n |n\rangle,\,n = 0, 1,\ldots;\;a \,|n\rangle = \sqrt{n}\,|n - 1\rangle, 
a^{\dagger}\, |n\rangle = \sqrt{n + 1}\,|n + 1\rangle\}$ provide the basis for the oscillator, whereas the eigenstates  
$\sigma_z |\pm 1\rangle = \pm \,|\pm 1\rangle$ span the space of the qubit. 
As the Hamiltonian (\ref{qOH}) is not known 
to be exactly solvable, numerous approximation schemes adapted to  various ranges of parameters have been advanced. For a
small detuning between the oscillator and the qubit frequencies, and also for a weak  qubit-oscillator coupling, the 
dynamical behavior of the interacting system is accurately described [\cite{JC1963}] by the rotating wave approximation. 
In another approach, the adiabatic approximation [\cite{IGMS2005}, \cite{AN2010}] scheme  is found to be appropriate in 
the large detuning limit ($\Delta \ll \omega$) as it utilizes the difference between the time scales of the slow-moving 
qubit and that of the fast-moving oscillator. Combining the virtues of these two approximation schemes a new procedure 
has been proposed [\cite{I2007}] that maintains a wide range of validity in the regime of  large values of both the coupling 
strength and the detuning of frequencies $(\lambda \sim O(\omega), \Delta \lesssim \omega )$. This generalization of the 
rotating wave approximation employs [\cite{I2007}], as a suitable intermediate step, a change of basis to that introduced 
by the adiabatic approximation [\cite{IGMS2005}, \cite{AN2010}]. Mimicking an inherent feature of the ordinary rotating wave 
approximation, the generalization advanced in [\cite{I2007}] retains, in the transformed basis, the `energy-conserving' 
one-particle transition elements in the Hamiltonian matrix which now assumes a direct sum of $2 \times 2$ diagonal blocks 
apart from an uncoupled ground state.  

\par

For the purpose of making our notations and subsequent applications clear, we, following [\cite{I2007}], now briefly review 
the diagonalization of the Hamiltonian in the generalized rotating wave approximation scheme, and explicitly list its 
eigenstates.  The adiabatic approximation physically signifies that the rapidly moving oscillator 
adjusts itself quickly to the slow changing qubit observable $\sigma_{z}$, which may be considered to
reside  in one of its eigenstates $|\pm1\rangle$. Setting $\Delta = 0$, the bipartite
Hamiltonian (\ref{qOH}) is readily diagonalized in the tensored basis $\ket{\pm 1,n_{\pm}}$, where 
 the  displaced number states read: $\ket{n_{\pm}}= \mathrm{D}^{\dagger}\left(\pm \frac{\lambda}{\omega}\right) \ket{n},
 \, \mathrm{D}\left(\alpha \right) = \exp\left(\alpha a^{\dagger}- \alpha^{*}a \right),\, \alpha \in \mathbb{C}$. 
 The eigenvalues of both the displaced basis states  $\ket{\pm 1,n_{\pm}}$ are degenerate: $E_{n} = 
 \omega\big(n - \frac{x}{4}\big), \, x=4\lambda^{2}/\omega^2$.
The  matrix elements for the projections of the displaced number states [\cite{IGMS2005}] are listed below:
\beq
\mathcal{M}_{m,n} \equiv \braket{m_{-}|n_{+}} =
 \begin{cases}
 (-1)^{m-n} \; {x}^{\frac{m-n}{2}} \; \exp\big(-\frac{x}{2}\big) \;
 \sqrt{n!/m!} 
\; L_n^{(m-n)}(x), & m \geq n \\
{x}^{\frac{n-m}{2}} \exp\big(-\frac{x}{2}\big) \;
 \sqrt{m!/n!} 
\; L_m^{(n-m)}(x) & m < n,
\label{Mmn}
\end{cases}
\eeq
where the associated Laguerre polynomial reads $L_{n}^{(j)}(x) = \sum_{k = 0}^{n} 
\,(-1)^{k}\, \binom{n + j}{n - k}\,\frac{x^{k}}{k!}$. The reality of the matrix element (\ref{Mmn}) leads to the identity 
\beq
\mathcal{M}_{m,n} =  (-1)^{m+n} \mathcal{M}_{n,m}
\label{IdenM}
\eeq
 that will be used later to simplify the density matrix elements.  
In the adiabatic approximation scheme, the degeneracy of the states are lifted by the qubit Hamiltonian that causes the mixing of the 
equi-displaced bipartite states $\ket{\pm 1,n_{\pm}}$. Diagonalizing the resulting $2 \times 2$ 
blocks associated with the same photon number the energy eigenvalues and the corresponding eigenstates are obtained [\cite{IGMS2005}]:
\beq
E^{(\pm)}_{n(\ge 0)} = 
\omega \Big( n- \frac{x}{4} \Big) \pm \dfrac{\widetilde{\Delta}}{2} L_{n}(x),\;\; \widetilde{\Delta} = \Delta \exp(-x/2),\;\; 
\ket{E^{(\pm)}_{n}} = \frac{1}{\sqrt{2}}(\ket{1, n_+} \pm \ket{-1, n_-}).
\label{E_adiabatic}
\eeq

\par

The basis obtained in (\ref{E_adiabatic}) is now utilized [\cite{I2007}] for presenting the generalized rotating 
wave approximation scheme. The Hamiltonian (\ref{qOH}), when expressed in the adiabatic eigenstates (\ref{E_adiabatic}), 
shows characteristics similar to the standard rotating wave approximation. Proceeding parallely, the tridiagonal matrix elements 
that contain the `energy conserving' one particle transitions are retained [\cite{I2007}], and the remote off-diagonal 
matrix elements that manifest higher order transitions 
are neglected within the generalized rotating wave approximation scheme. The ground state is now decoupled from all 
other states, whereas the remaining matrix elements are organized in a tower of  $2 \times 2$ diagonal blocks  
[\cite{I2007}] as
\beq
\begin{pmatrix}
E^{(+)}_{n-1} & \zeta_{n}  \\
\zeta_{n} & E^{(-)}_{n}
\end{pmatrix},\qquad
\zeta_{n} = \dfrac{\Delta}{2} \; \mathcal{M}_{n-1,n} =\dfrac{\widetilde{\Delta}}{2}\,
\sqrt{\frac{x}{n}} \,L^{(1)}_{n-1}(x), \; n \ge 1.
\label{2X2GRWA}
\eeq 
 The energies of the singlet ground state, and the infinite tower of the doublets read [\cite{I2007}]
 \bea
\mathcal{E}_{0} & \equiv & E^{(-)}_{0} =-\frac{\omega x}{4}-\frac{\widetilde{\Delta}}{2},\nn\\ 
\mathcal{E}^{(\pm)}_{n (\ge 1)}& = &\omega \Big(n - \frac{1}{2} - \frac{x}{4}\Big) 
 + \frac{\widetilde{\Delta}}{4}
\Big(L_{n-1} (x) - L_{n}(x)\Big) \nn \\
& \pm &\frac{1}{2}\sqrt{ \Big \lgroup  \omega -\frac{\widetilde{\Delta}}{2} 
\Big( L_{n-1} (x) + L_{n} (x) \Big) \Big \rgroup^2
+  x \,\frac{\widetilde{\Delta}^2 }{n} \Big \lgroup L_{n-1}^{(1)}(x) \Big \rgroup^2}. 
\label{energy_GRWA}
\eea
The corresponding eigenstates are explicitly given by
\beq
|{\mathcal E}_0 \rangle \equiv \ket{{E}_0^{(-)}} = \dfrac{1}{\sqrt{2}} 
\Big( \ket{1,0_{+}} - \ket{-1,0_{-}} \Big),\;\;
|{\mathcal E}_{n (\ge 1)}^{(\pm)} \rangle = \mu^{(\pm)}_{n}  | E^{(+)}_{{n-1}} 
\rangle \pm \dfrac{\zeta_n}{|\zeta_n|} \mu^{(\mp)}_{n}|E^{(-)}_{n} \rangle, 
\label{Hn_eigenstate}
\eeq
where we  abbreviate: $\chi_n=\sqrt{\zeta_n^2+\varepsilon_{n}^2},\,\varepsilon_{n} 
=\frac{E^{(+)}_{n-1}-E^{(-)}_{n}}{2},\, \mu^{(\pm)}_{n} = 
\sqrt{\frac{\chi_n \pm \varepsilon_{n}} {2\chi_n}}$. The completeness requirement of the bipartite basis states 
(\ref{Hn_eigenstate}) now reads: 
\beq
|{\mathcal E}_{0} \rangle \langle {\mathcal E}_{0}| + \sum_{n = 1}^{\infty} 
\left(|{\mathcal E}_{n}^{(+)} \rangle \langle {\mathcal E}_{n}^{(+)}| +  |{\mathcal E}_{n}^{(-)} \rangle 
\langle {\mathcal E}_{n}^{(-)}|\right) = \sum_{n = 0}^{\infty} 
\left(| E_{n}^{(+)} \rangle \langle  E_{n}^{(+)}| +  |E_{n}^{(-)} \rangle 
\langle E_{n}^{(-)}|\right) = \mathbb{I}
\label{completeness}
\eeq  

\par

With the above construction of the energy eigenstates via the generalized rotating wave approximation procedure in 
hand, we now study the time evolution of the entanglement of the bipartite system.
The initial qubit-oscillator hybrid entangled states read
\beq 
\ket{\Psi(0)}_{(\pm)}=\dfrac{1}{2} \left \lgroup \ket{1} \Big( \ket{\alpha} + \ket{-\alpha} \Big)
\pm \ket{-1} \Big( \ket{\alpha} - \ket{-\alpha} \Big) \right \rgroup, 
\qquad  \ket{\alpha} = \mathrm{D}(\alpha)\,\ket{0}.
\label{t0quasiBell}
\eeq
As the Schr\"{o}dinger cat states $\sim\, \ket{\alpha} \pm \ket{-\alpha}$ are mutually orthogonal, the initial states (\ref{t0quasiBell})
may be regarded as examples of the hybrid Bell states.
Even though our subsequent analysis may be developed with both the initial states (\ref{t0quasiBell}), we, for the purpose 
of notational simplicity, quote the results for the choice
$\ket{\Psi(0)}_{(-)}$ and omit the subscript hereafter. The complete basis states 
(\ref{Hn_eigenstate}) immediately implement the mode expansion
\beq
\ket{\Psi(0)} = \mathcal{C}_{0} \ket{\mathcal{E}_{0}} 
+ \sum_{n=1}^{\infty} \mathcal{C}_{n}^{(\pm)} \ket{\mathcal{E}_{n}^{(\pm)}}.
\label{InExpansion}
\eeq
The coefficients of the vector space expansion (\ref{InExpansion}) are listed below:
\bea
\mathcal{C}_{0} &=& \frac{1}{\sqrt{2}}\exp \left \lgroup  
-\frac{|\alpha_{-}|^{2}}{2} + i\phi_{\alpha} \right \rgroup,\;\;\phi_{\alpha} 
= \frac{\lambda}{\omega} \,\mathrm{Im}(\alpha), \nn\\
\mathcal{C}_{n}^{(\pm)} &=& \frac{1}{\sqrt{2}} \left \lgroup \exp \left \lgroup -
\frac{|\alpha_{+}|^{2}}{2} -i\phi_{\alpha} \right \rgroup
 P^{(+)}_{n-1} \dfrac{\alpha_{+}^{n-1}}{\sqrt{(n-1)!}} 
\left \lgroup  \mu^{(\pm)}_{n} \pm 
\frac{\alpha_{+}}{\sqrt{n}} \frac{\zeta_n}{|\zeta_n|} \mu^{(\mp)}_{n}  
\right \rgroup \right. \nn   \\
& & \left. -  \exp \left \lgroup  
-\frac{|\alpha_{-}|^{2}}{2} + i\phi_{\alpha} \right \rgroup 
 P^{(-)}_{n-1} \frac{\alpha_{-}^{n-1}}{\sqrt{(n-1)!}} 
\left \lgroup \mu^{(\pm)}_{n} \mp 
\frac{\alpha_{-}}{\sqrt{n}} \frac{\zeta_n}{|\zeta_n|} \mu^{(\mp)}_{n} 
\right\rgroup\right \rgroup,
\label{C_def}
\eea
where the following definitions are used:
$\alpha_{\pm} = \alpha \pm \frac{\sqrt{x}}{2}, P_{n}^{(\pm)} = \frac{1 \pm (-1)^{n}}{2}$. The basis of approximate eigenstates 
(\ref{Hn_eigenstate}) of the Hamiltonian (\ref{qOH}) readily yields the
time evolution of the initial state (\ref{InExpansion}):
\beq
\ket{\Psi(t)} = \mathcal{C}_{0}(t) \ket{\mathcal{E}_{0}} 
+ \sum_{n=1}^{\infty} \mathcal{C}_{n}^{(\pm)}(t) \ket{\mathcal{E}_{n}^{(\pm)}}, 
\label{t_BipState}
\eeq 
where the coefficients read: $\mathcal{C}_{0}(t) = 
\mathcal{C}_{0} \exp\left(-i\mathcal{E}_{0} t\right),\,
\mathcal{C}_{n}^{(\pm)}(t)=\mathcal{C}_{n}^{(\pm)} \exp\Big(-i\mathcal{E}^{(\pm)}_{n} t\Big)$. The time dependent density 
matrix of the bipartite pure state has the tensored form
\beq
\rho(t) = \ket{\Psi(t)} \! \bra{\Psi(t)}.
\label{t_DenMa} 
\eeq

\par

The reduced density matrices of the individual subsystems are extracted by partial tracing over the 
complementary subspaces in the full Hilbert space.  For instance, the  tracing over the oscillator states   yields the 
reduced density matrix of the qubit: 
\beq
\rho_{\cal Q} \equiv \hbox{Tr}_{\cal{O}} \, \rho = 
\begin{pmatrix}
\frac{1}{2} + \varrho  &   \xi \\
\xi^{*} &  \frac{1}{2} - \varrho
\end{pmatrix},
\label{Qubit_DenMa}
\eeq
where the matrix elements read
\bea
\varrho &=&  \mathrm{Re} \Big \lgroup {\mathcal{C}_{0}(t) \mathcal{A}_{1}(t)}^{*}  
+ \sum_{n=1}^{\infty} {\mathcal{A}_{n+1}(t)}^{*} \mathcal{B}_{n}(t) \Big \rgroup,\nn\\
\xi &=& - \frac{1}{2}|\mathcal{C}_{0}(t)|^{2} \mathcal{M}_{0,0}- \frac{1}{2} 
\sum_{n=1}^{\infty} \Big \lgroup \Big( {\mathcal{C}_{0}(t)}^{*} \mathcal{A}_{n}(t) + 
(-1)^{n} \mathcal{C}_{0}(t) {\mathcal{A}_{n}(t)}^{*}  \Big) \mathcal{M}_{0,n-1}\nn \\
& &+ \Big( \!{\mathcal{C}_{0}(t)}^{*} \mathcal{B}_{n}(t)\! +\!
(-1)^{n} \mathcal{C}_{0}(t) {\mathcal{B}_{n}(t)}^{*} \! \Big) \!\mathcal{M}_{0,n} \Big\rgroup\! 
\! \!+\! \frac{1}{2}\! \sum_{n,m=1}^{\infty}\!\!\! \Big \lgroup {\!\mathcal{A}_{n}(t)}^{*} 
\mathcal{A}_{m}(t)\mathcal{M}_{n-1,m-1} \nn \\
& & - {\mathcal{B}_{n}(t)}^{*} 
\mathcal{B}_{m}(t) \mathcal{M}_{n,m} + \Big({\mathcal{A}_{n}(t)}^{*}
\mathcal{B}_{m}(t) + (-1)^{m+n} \mathcal{A}_{n}(t) 
{\mathcal{B}_{m}(t)}^{*}\Big) \mathcal{M}_{n-1,m}  
\Big \rgroup,\nn\\
\mathcal{A}_{n}(t) &=& \mu_{n}^{(+)} \mathcal{C}_{n}^{(+)}(t) + \mu_{n}^{(-)} \mathcal{C}_{n}^{(-)}(t),\quad
\mathcal{B}_{n}(t) = \frac{\zeta_{n}}{|\zeta_{n}|} 
\left\lgroup \mu_{n}^{(-)} \mathcal{C}_{n}^{(+)}(t) - \mu_{n}^{(+)} \mathcal{C}_{n}^{(-)}(t) \right\rgroup.
\label{2X2_DenMaElts}
\eea
In the above expression we have used the identity (\ref{IdenM}).
The expectation value of the qubit spin variable $\langle\sigma_{z}\rangle$ measures the statistical average of the population 
inversion. For the density matrix 
(\ref{Qubit_DenMa}) it is given by
\beq
\langle\sigma_{z}\rangle \equiv \mathrm{Tr}(\sigma_{z}\rho_{\mathcal{Q}}) = 2 \varrho.
\label{PopInv}
\eeq
In general, the pair of eigenvalues of the qubit density matrix (\ref{Qubit_DenMa})  
\beq
\frac{1}{2} \pm \varpi,\qquad \varpi= 
\sqrt{\varrho^2+|\xi|^2}
\label{rho_eigenvalue}
\eeq
allows us to compute its von Neumann entropy  $S(\rho_{\cal Q}) \equiv - \hbox {Tr} (\rho_{\cal Q} \, \log \rho_{\cal Q})$ as
\beq
S_{\cal Q} = - \left(\frac{1}{2} + \varpi\right) \log \left(\frac{1}{2} + \varpi\right) - \left(\frac{1}{2} - 
\varpi\right) \log \left(\frac{1}{2} - \varpi\right).
\label{entropy_defn}
\eeq

\par

The bipartite density matrix (\ref{t_DenMa}) also similarly produces the reduced density matrix of the oscillator  via 
partial tracing on the qubit Hilbert space: $\rho_{\cal{O}} \equiv \hbox{Tr}_{\cal Q} \, \rho$. Its explicit construction is 
listed below:
\bea
\rho_{\mathcal{O}}(t) \! \! \! &=& \! \! \! |\mathcal{C}_{0}(t)|^{2} P_{0,0}^{(+)} \! + \! 
\sum_{n=1}^{\infty} \Big( \mathcal{C}_{0}(t) {\mathcal{A}_{n}(t)}^{*} P_{0,n-1}^{(-)}  \! + \!
{\mathcal{C}_{0}(t)}^{*} \mathcal{A}_{n}(t) P_{n-1,0}^{(-)} 
+ \mathcal{C}_{0}(t) {\mathcal{B}_{n}(t)}^{*} P_{0,n}^{(+)} \; \; \; \nn \\ 
& &+ {\mathcal{C}_{0}(t)}^{*} \mathcal{B}_{n}(t) P_{n,0}^{(+)} \Big) +
\sum_{n,m=1}^{\infty} \Big( \mathcal{A}_{n}(t) {\mathcal{A}_{m}(t)}^{*} P_{n-1,m-1}^{(+)} 
+ \mathcal{B}_{n}(t) {\mathcal{B}_{m}(t)}^{*} P_{n,m}^{(+)} \nn \\
& &+ \mathcal{B}_{n}(t) {\mathcal{A}_{m}(t)}^{*} P_{n,m-1}^{(-)} +
\mathcal{A}_{n}(t) {\mathcal{B}_{m}(t)}^{*} P_{n-1,m}^{(-)} \Big), 
\label{density_oscillator}
\eea
where the symmetrized projection operators read
$P_{n,m}^{(\pm)}= \frac{1}{2} \left( \ket{n_{+}}\bra{m_{+}} \pm \ket{n_{-}}\bra{m_{-}} \right),\,(n, m = 0, 1, \ldots)$. The 
density matrix (\ref{density_oscillator}) obeys the normalization condition: 
$\mathrm{Tr}\,\rho_{\mathcal{O}}(t) = 1$. For later use we quote the initial oscillator density matrix: 
$\rho_{\mathcal{O}}(0) =\frac{1}{2}\,(\ket{\alpha}\bra{\alpha} + \ket{-\alpha}\bra{-\alpha})$.

\par

To obtain measures of the entanglement and the mixedness of the  bipartite system  the von Neumann entropy of the reduced 
density matrix of the qubit  may be considered.  It is well-known [\cite{AL1970}] that
if  a composite system, comprising of two subsystems, resides in a pure state, the entropies of both subsystems are equal. 
In the present example it holds for the oscillator with an infinite dimensional Hilbert space, and the  two-level qubit 
interacting with it: $S_{\cal Q} = S_{\cal O} \equiv S$.
\section{Quasi-probability distributions on the oscillator phase space}
\setcounter{equation}{0}
\subsection{The diagonal Sudarshan-Glauber $P$-representation}
\label{Sudarshan}

It is well-known that the coherent state representation is overcomplete [\cite{Schleich2001}]. 
Employing the overcompleteness property, it has been established  
[\cite{S1963}, \cite{G1963}] that an arbitrary oscillator density matrix $\rho_{\cal O}$ may be realized  via  diagonal coherent state 
projectors:
\beq
\rho_{\cal O} = \int P(\beta,\beta^{*}) \ket{\beta}  \bra{\beta}  \mathrm{d}^{2}\beta.
\label{P_def}
\eeq
The normalization of the  density matrix $\rho_{\cal O}$ ensures the property
\beq 
\int P(\beta, \beta^{*}) \mathrm{d}^{2}\beta = 1, 
\label{P_norm}
\eeq
whereas its Hermiticity implies  $P(\beta, \beta^{*})$ to be real. The  $P$-representation, however,  is a general 
distribution of indeterminate sign as  it possibly includes highly 
singular derivatives of the $\delta$-function [\cite{S1963}]. For the coherent state, which corresponds as closely as 
possible to a classical harmonic oscillator state with a complex amplitude, the $P$-representation is a positive definite 
$\delta$-function [\cite{Schleich2001}] which is a legitimate measure of the classical probability density. For this reason 
the coherent state is considered to be classical. On the other hand, if the diagonal kernel  $P(\beta,\beta^{*})$ can 
not be considered a valid probability measure due to its negativity, the oscillator state given by the density matrix 
$\rho_{\cal O}$ exhibits nonclassical features. The diagonal $P$-representation is known to be unique, even though it
may be realized in different equivalent forms [\cite{M2009}]. Towards inverting the diagonal representation  (\ref{P_def})
the Fourier transform is employed [\cite{Schleich2001}]:
\beq
P(\beta,\beta^{*})= \dfrac{\exp(|\beta|^{2})}{\pi ^{2}} 
\int \bra{-\gamma}\rho_{\cal{O}}\ket{\gamma} \,\exp(|\gamma |^{2})\, 
\exp(\beta  \gamma^{*} - \beta^{*} \gamma)\, \mathrm{d}^{2}\gamma.
\label{P_evaluation}
\eeq
The invertibility of the relations (\ref{P_def}, \ref{P_evaluation}) suggests that both 
 the density matrix $\rho_{\cal{O}}$ and the diagonal representation $P(\beta,\beta^{*})$ encapsulate equivalent informations.
We now explicitly evaluate the pseudo probability distribution corresponding to the oscillator density matrix 
(\ref{density_oscillator}): 
\bea
P(\beta,\beta^{*}) &=&  \frac{1}{2} |\mathcal{C}_{0}(t)|^{2}  D^{(+)}_{0,0}  
+ \mathrm{Re} \Big \lgroup   \mathcal{C}_{0}(t)^{*} 
\sum_{n=1}^{\infty}  \Big( \mathcal{A}_{n}(t) D^{(-)}_{0,n-1} 
+  \mathcal{B}_{n}(t) D^{(+)}_{0,n} \Big) \nn \\
& &+ \sum_{n,m=1}^{\infty} \mathcal{B}_{n}(t)^{*} \mathcal{A}_{m}(t) D^{(-)}_{n,m-1} \Big \rgroup 
+ \dfrac{1}{2} \sum_{n,m=1}^{\infty} \Big( \mathcal{A}_{n}(t)^{*} \mathcal{A}_{m}(t) D^{(+)}_{n-1,m-1} \nn \\
& &+  \mathcal{B}_{n}(t)^{*} \mathcal{B}_{m}(t)  D^{(+)}_{n,m} \Big),
\label{P_function}
\eea
where the arbitrarily singular derivatives of the $\delta$-functions read
\bea
D_{n,m}^{(\pm)} &=& \frac{1}{\sqrt{n!m!}} \Big \lgroup \exp(|\beta_{+}|^{2}) 
\Big( -\frac{\partial}{\partial \beta_{+}^{*}}\Big)^{n} 
\Big( -\frac{\partial}{\partial \beta_{+}}\Big)^{m} \delta^{(2)}(\beta_{+})\nn \\
& &\pm \exp(|\beta_{-}|^{2})  \Big( -\dfrac{\partial}{\partial \beta_{-}^{*}}\Big)^{n} 
\Big( -\frac{\partial}{\partial \beta_{-}}\Big)^{m} \delta^{(2)}(\beta_{-}) \Big \rgroup, 
\quad \beta_{\pm}=\beta \pm \frac{\sqrt{x}}{2}.
 \label{D_singular}
\eea
The diagonal $P$-representation (\ref{P_function}) satisfies the normalization condition (\ref{P_norm}), and is obviously 
real. At $t=0$, the initial hybrid entangled state (\ref{t0quasiBell}) gives rise to a positive definite 
$P$-representation: $P(\beta,\beta^{*})|_{{t=0}} = 2^{-1}\, \exp(-(|\alpha|^{2}-|\beta|^{2}))\,
(\delta^{(2)}(\alpha-\beta) + \delta^{(2)}(\alpha+\beta))$. At an arbitrary time, however, a large number of 
interaction-dependent modes develop, and produce quantum interferences among themselves. The quantum properties in
(\ref{P_function}) are evident due to the appearance of rapidly oscillating singular derivatives of $\delta$-function in the 
$P$-representation, which at time $t > 0$ is clearly not a nonnegative definite quantity. The existence of negativity of the  diagonal 
$P$-representation (\ref{P_function}) points towards the manifestations of specific quantum effects such as anti-bunching [\cite{M1982}] 
 and quadrature squeezing [\cite{Yurke1985}]. Other phase space pseudo probability functions such as Wigner $W$-distribution and the
 Husimi $Q$-function may be obtained from the diagonal $P$-representation after its suitable smoothing by the Gaussian kernels.                                                                                                                                                                                                                                                                                                                                                                                                                                                                                                                                                                                                                                                                                                                                                                                                                                                                                                                                                                                                                                                                                                                                                                                                                                    
\subsection{The Wigner $W$-distribution}
\label{Wigner}
The Wigner phase space distribution for a quantum state given by the oscillator density matrix (\ref{density_oscillator}) 
is described  [\cite{Schleich2001}] as
\beq
W(\beta, \beta^{*}) = \dfrac{1}{\pi ^{2}} \int \mathrm{Tr}(\mathrm{D}(\gamma)\rho_{\cal{O}}) \exp\Big( \beta \gamma^{*}
- \beta^{*} \gamma \Big) \mathrm{d}^{2}\gamma,
\label{W_def}
\eeq
which is a real-valued function maintaining the normalization property: 
\beq
\int W(\beta,\beta^{*})\; \mathrm{d}^{2}\beta = 1. 
\label{W_normalization}
\eeq
 A characteristic of the Wigner distribution that distinguishes it from a classical probability density, is that its 
 integral over a given subregion of the phase space may be negative or greater than one. 
As an efficient computational algorithm the $W$-distribution has been expressed [\cite{MK1993}] as a sum of an infinite 
series of the expectation values of the oscillator density matrix $\rho^{\phantom{A}}_{\cal{O}}$ in the displaced number 
basis states:
\beq
W (\beta, \beta^{*}) = \dfrac{2}{\pi} \sum_{k=0}^{\infty} (-1)^{k} \bra{\beta,k} 
\rho^{\phantom{A}}_{\cal{O}} \ket{\beta,k}, \qquad \ket{\alpha, n} = \mathrm{D}(\alpha) \ket{n}.
\label{W_series}
\eeq
On the other hand, the convolution of a smoothing Gaussian function of variance $1/2$ on the phase space with the diagonal  
$P$-representation leads to the  $W$-distribution that is free from the divergences brought forward by the derivatives of 
the $\delta$-function contained in the expression (\ref{P_function}) of the $P$-representation:
\beq
W(\beta,\beta^{*}) = \frac{2}{\pi}\int P(\gamma, \gamma^{*}) \exp(-2|\beta-\gamma|^{2}) 
\;\mathrm{d}^{2}\gamma.
\label{P_int_W}
\eeq 
Employing the explicit evaluation (\ref{P_function}) of the $P$-representation and the convolution integral 
(\ref{P_int_W}) we now provide a direct derivation of the $W$-distribution. Towards this we employ the identity 
\begin{align}
\int &\exp \Big( -2|\beta - \gamma|^{2}+|\gamma |^{2}\Big) 
\Big( -\dfrac{\partial}{\partial \gamma^{*}} \Big)^{n} 
\Big( -\dfrac{\partial}{\partial \gamma} \Big)^{m} \delta^{(2)}(\gamma)\, \mathrm{d}^{2}\gamma\nn\\
&= 2^{n+m} \beta^{n} \beta^{*m}  \exp(-2|\beta|^{2})\,{}_2F_0\Big( \! \! -n,-m;\phantom{}_{-} ;
-\frac{1}{4|\beta|^{2}}\Big),
\label{Int_hyper}
\end{align} 
where the  hypergeometric sum reads $ \displaystyle{ {}_2F_0(\mathsf{x},\mathsf{y};\phantom{}_{-}; \tau)
= \sum_{k=0}^{\infty} (\mathsf {x})_{k} (\mathsf{y})_{k}\, \frac{\tau^{k}}{k!},\,
(\mathsf {x})_{k} = \prod_{\ell = 0}^{k - 1}(\mathsf {x} + \ell) }$.  
The above hypergeometric function with the negative integral numerator may be  expressed [\cite{M1939}] via the Charlier 
polynomial: $\mathrm{c}_{k} (\ell; \tau) = {}_{2}F_0\left(-k,-\ell;\phantom{}_{-} ;
-\frac{1}{\tau}\right)\,\forall \tau > 0$. To emphasize the equal footing of the integers $(k, \ell)$ 
in (\ref{Int_hyper}) we, however, do not use this notation here. The integral representation 
(\ref{P_int_W}) aided by the identity (\ref{Int_hyper}) now generate the $W$-distribution:
\bea
W(\beta,\beta^{*}) \! \! \! \! &=& \! \! \! \! |\mathcal{C}_{0}(t)|^{2} 
\; \mathcal{H}_{0,0}^{(+)} + \!2\; \mathrm{Re} \! \Big \lgroup \mathcal{C}_{0}(t)^{*}  
\sum_{n=1}^{\infty} \! \!  \Big( \! 
\mathcal{A}_{n}(t) \mathcal{H}_{0,n-1}^{(-)}(\beta,\beta^{*}) \! + \! 
\mathcal{B}_{n}(t) \mathcal{H}_{0,n}^{(+)}(\beta,\beta^{*}) \! \Big)  \nn \\
 & +& \! \! \! \! \! \! \! \!
\sum_{n,m=1}^{\infty} \! \! \!  \mathcal{B}_{n}(t)^{*} \mathcal{A}_{m}(t) 
\mathcal{H}_{n,m-1}^{(-)} (\beta,\beta^{*}) \Big \rgroup   \! \! + \! \! \! \!
\sum_{n,m=1}^{\infty} \! \! \! \Big( \mathcal{A}_{n}(t)^{*} \mathcal{A}_{m}(t) 
\mathcal{H}_{n-1,m-1}^{(+)} (\beta,\beta^{*}) \nn \\
 &+& \! \! \! \!
\mathcal{B}_{n}(t)^{*} \mathcal{B}_{m}(t) \mathcal{H}_{n,m}^{(+)} (\beta,\beta^{*})
\Big),
\label{wigner_function}
\eea
where the weights involving the Gaussian functions are given by
\bea
\mathcal{H}^{(\pm)}_{n,m}(\beta, \beta^{*}) &=& \dfrac{2^{n+m}}{\pi \sqrt{n!m!}} 
\Big \lgroup \beta_{+}^{n} \beta_{+}^{*^m} \exp(-2|\beta_{+}|^{2}) \;
{}_2F_0\Big( -n,-m;\phantom{}_{-} ; -\dfrac{1}{4|\beta_{+}|^{2}}\Big)  \nn \\ 
& &\pm \beta_{-}^{n} \beta_{-}^{*^m}  \exp(-2|\beta_{-}|^{2}) \;
{}_2F_0\Big( -n,-m;\phantom{}_{-} ; -\dfrac{1}{4|\beta_{-}|^{2}}\Big) \Big \rgroup.
\label{HW}
\eea

\par

An alternate derivation of the Wigner $W$-distribution follows from sum rule (\ref{W_series}). Towards this we list the 
matrix elements
\beq
\braket{n_{\pm}|\alpha,k} = (-1)^{k}
 \frac{\alpha_{\pm}^{n}\alpha_{\pm}^{*^k}}{\sqrt{n! k!}}\,
 \exp\Big(-\frac{|\alpha_{\pm}|^{2}}{2} \mp i \phi_{\alpha}\Big) \;
{}_2F_0\Big( -n,-k;\phantom{}_{-}; -\frac{1}{|\alpha_{\pm}|^{2}}\Big)
\label{NAlphaK}
\eeq
that facilitate the evaluation of the series (\ref{W_series}). Another necessary tool for the present derivation is the 
 identity
\begin{align}
\sum_{k=0}^{\infty} \frac{(-1)^{k} \tau^{k}}{k!}\, {}_2F_0\Big( -n,-k;\phantom{}_{-} ; -\frac{1}{\tau}\Big) &
\,{}_2F_0\Big( -k,-m;\phantom{}_{-} ; -\dfrac{1}{\tau}\Big) = \nn \\
& 2^{n+m} \exp(-\tau) \; {}_2F_0\Big( -n,-m;\phantom{}_{-} ; -\frac{1}{4\tau}\Big)
\label{IdentityHyper}
\end{align}
that readily follows  [\cite{M1939}] from the bilinear generating function of the Charlier polynomials. The series sum 
(\ref{W_series}), in conjunction with the results (\ref{NAlphaK}, \ref{IdentityHyper}), 
now reproduce the $W$-distribution obtained in (\ref{wigner_function}). This serves as a consistency check on our 
derivations (\ref{P_function}, \ref{wigner_function}) of the phase space distributions in the present model.

\par

The negativity of the Wigner function is an indication towards the existence of nonclassical properties of the states. 
At $t = 0$ the $W$-distribution for our initial hybrid entangled state 
(\ref{t0quasiBell}) is positive definite: $W(\beta,\beta^{*})|_{t=0} = \pi^{-1}\,
(\exp(-2|\alpha - \beta|^{2}) + \exp(-2|\alpha + \beta|^{2}))$. At a later  time, however, interactions produce a large 
number of modes that  interfere with each other. Interference fringes gives rise to appearance of phase space domains 
representing negative values of the 
$W$-distribution. A fruitful measure of negativity relate to the volume of the negative part of the Wigner function on 
the phase space [\cite{KZ2004}]:
\beq 
\delta_{W} = \int |W(\beta)| \mathrm{d^2}\beta-1.
\label{negativity}
\eeq
We will return to the topic in Sec. \ref{DelocalQuantum}.

\subsection{The Husimi $Q$-function}
\label{Husimi} 

The Husimi $Q$-function [\cite{Schleich2001}] is a quasi probability distribution  defined  as expectation value of the oscillator density matrix 
in an arbitrary coherent state. It assumes nonnegative values on the phase space in contrast to the other phase space quasi probabilities. Being easily 
computable it is extensively employed [\cite{SA2002}, \cite{IWAH2002}] in the study of the 
occupation on the phase space. 
For our reduced density matrix of the oscillator (\ref{density_oscillator}) the  $Q$-function 
\beq
Q(\beta,\beta^{*}) = \frac{1}{\pi} \bra\beta\rho_{\cal O}\ket\beta
\label{Q_defn}
\eeq 
 maintains the normalization restriction: $\int Q(\beta, \beta^{*}) \mathrm{d}^{2}\beta = 1$ and the bounds: 
 $0 \leq Q(\beta,\beta^{*}) \leq \frac{1}{\pi}$. Our
 construction of the oscillator density matrix (\ref{density_oscillator}) now yields the time-evolution of the $Q$-function:
\bea
Q(\beta,\beta^{*}) &=& \dfrac{1}{2}  |\mathcal{C}_{0}(t)|^{2} H^{(+)}_{0,0}(\beta,\beta^{*})
 + \mathrm{Re} \Big \lgroup \mathcal{C}_{0}(t)^{*} 
\sum_{n=1}^{\infty} \Big( \mathcal{A}_{n}(t) H^{(-)}_{0,n-1}(\beta,\beta^{*})\nn  \\
&+& \mathcal{B}_{n}(t) H^{(+)}_{0,n}(\beta,\beta^{*}) \Big) + 
\sum_{n,m=1}^{\infty} \! \mathcal{A}_{n}(t)^{*} 
\mathcal{B}_{m}(t) H^{(-)}_{n-1,m}(\beta,\beta^{*}) \Big \rgroup \nn \\
&+&\dfrac{1}{2}\! \sum_{n,m=1}^{\infty} \!\Big( \mathcal{A}_{n}(t)^{*} 
\mathcal{A}_{m}(t) H^{(+)}_{n-1,m-1}(\beta,\beta^{*}) 
+\mathcal{B}_{n}(t)^{*} \mathcal{B}_{m}(t) H^{(+)}_{n,m}(\beta,\beta^{*}) \Big),
 \label{Q_function}
\eea
where the weight functions on the phase space read
\beq
H^{(\pm)}_{n,m} (\beta, \beta^{^*})= \dfrac{1}{\pi \sqrt{n!m!}} 
\Big( \beta_{+}^{n} \beta_{+}^{*^m} \exp(-|\beta_{+}|^{2})  \pm 
\beta_{-}^{n} \beta_{-}^{*^m}  \exp(-|\beta_{-}|^{2}) \Big).
\label{Q_weight}
\eeq
A convolution integral of the diagonal $P$-representation with a Gaussian weight of unit variance
produce [\cite{Schleich2001}] the well-behaved nonnegative Husimi $Q$-function as follows:
\beq
Q(\beta,\beta^{*}) = \frac{1}{\pi}\int P(\gamma,\gamma^{*})  \exp(-|\beta-\gamma|^{2})  
\;\mathrm{d}^{2}\gamma.
\label{P_int_Q}
\eeq
The explicit constructions of the $P$-representation (\ref{P_function}) and that of the $Q$-function 
(\ref{Q_function}) obey the consistency check (\ref{P_int_Q}). Another convolution property that express the $Q$-function 
via the $W$-distribution reads [\cite{Schleich2001}]:
\beq
Q(\beta,\beta^{*})=\dfrac{2}{\pi}\int W(\gamma,\gamma^{*}) 
\exp(-2|\beta-\gamma|^{2}) \; \mathrm{d}^{2}\gamma.
\label{Q_int_W}
\eeq
Employing the integral representation (\ref{Int_hyper}) it follows that the phase space 
quasi probability densities (\ref{wigner_function}, \ref{Q_function}) maintain the integral sum rule
(\ref{Q_int_W}). The convolution relations (\ref{P_int_W}, \ref{Q_int_W}) suggest an interesting feature. The diagonal 
$P$-representation containing singular derivatives of $\delta$-functions is smoothed at two stages by a Gaussian weight 
of variance $1/2$. After the first smoothing operation 
the $W$-distribution, which is  nonsingular but not non-negative, appears. Subsequent smoothing operation produces a 
nonsingular \textit{and} nonnegative Husimi $Q$-function. If a single smoothing operation (\ref{P_int_Q}) is performed 
on $P$-representation  via the Gaussian weight of variance $1$, the $Q$-function is directly reproduced. We symbolically 
summarize this as 
$[[P]_{1/2}]_{1/2} = [P]_{1}$, where the notation reads $[\mathcal{X}]_{s} =  
\frac{1}{\pi\, s}\int \mathcal{X}(\gamma,\gamma^{*})  
\exp\left(-\frac{|\beta-\gamma|^{2}}{s}\right)\,\mathrm{d}^{2}\gamma$.     

\par
 
One of the utilities of the $Q$-function is that it provides a convenient evaluation of
the expectation values of the operators expressed in their antinormal ordered form [\cite{Schleich2001}]. The
first and second moments of the quadrature variable defined as
\beq
X_{\theta}=\frac{1}{2} \left(a \exp(-i\theta)+ a^{\dagger} \exp(i\theta)\right)
\label{angle_hermitian_operator}
\eeq 
may be expressed [\cite{Schleich2001}] via the following phase space integrals over the complex plane:
\bea
\langle X_{\theta} \rangle &\equiv& \hbox {Tr}\,(X_{\theta} \rho_{\cal O}(t)) = \frac{1}{2} \int 
(\beta e^{-i\theta}+ \beta^* e^{i\theta}) Q(\beta, \beta^{*}) d^2 \beta,\nn\\
\langle X_{\theta}^2 \rangle &\equiv& \hbox {Tr}\,(X_{\theta}^2 \rho_{\cal O}(t)) = \frac{1}{4} \int
\Big( (\beta e^{-i\theta}+ \beta^* e^{i\theta})^2-1 \Big ) Q(\beta, \beta^{*}) d^2 \beta.
\label{moment_def}
\eea
The $Q$-function (\ref{Q_function}) now allows explicit determination of the above quadrature moments: 
\bea
\langle X_{\theta} \rangle  &=& \mathrm{Re}\Big(\mathbf{G}_{1}(t)\exp(-i\theta)\Big) -
\sqrt{x}\varrho \cos\theta, 
\label{1stMoment} \\
\langle X_{\theta}^2 \rangle &=& \frac{1}{2}\, 
\mathrm{Re} \Big(\mathbf{G}_{2}(t)\exp(-i2\theta) - 
\sqrt{x} \; \mathbf{F}_{1}(t) \; (1+\exp(-i2\theta))\Big) \nn \\ 
& & +  \frac{1}{2} \Big( \mathbf{N}_{1}(t) - \dfrac{1}{2} \Big)
 \! + \! \dfrac{x}{4}\cos^{2}\theta
\label{2ndMoment} 
\eea
where the time-dependent coefficients 
$\{\mathbf{G}_{k}(t), \mathbf{N}_{k}(t), \mathbf{F}_{k}(t)|k\in (1, 2)\}$ expressed as sums over Fourier modes read:
\bea
\mathbf{G}_{k}(t) \! \! \! \! &=& \! \! \! \! \sqrt{k} \; \mathcal{C}_{0}(t)^{*} \mathcal{B}_{k}(t) \!
+ \! \! \sum_{n=1}^{\infty} \! \! \Big \lgroup \! \! \sqrt{(n)_{k}} \; \mathcal{A}_{n}(t)^{*} 
\mathcal{A}_{n+k}(t)\nn \\
& & +\sqrt{(n+1)_{k}} \; \mathcal{B}_{n}(t)^{*} \mathcal{B}_{n+k}(t) \! \Big \rgroup, \nn\\
\mathbf{N}_{k}(t) \! \! \! &=& \! \! \! |\mathcal{C}_{0}(t)|^{2}+ 
\sum_{n=1}^{\infty} \Big \lgroup n^{k} \; \mathcal{A}_{n}(t)^{*} \mathcal{A}_{n}(t)
+ (n+1)^{k} \; \mathcal{B}_{n}(t)^{*} \mathcal{B}_{n}(t) \Big \rgroup, \nn \\
\mathbf{F}_{k}(t)  \! \! \! &=& \! \! \! k \; \mathcal{C}_{0}(t)^{*} \mathcal{A}_{2}(t) +
\sum_{n=1}^{\infty} \Big \lgroup \sqrt{n} \;(n+1)_{k-1} \mathcal{A}_{n}(t)^{*} \mathcal{B}_{n}(t)\nn \\
& & + \sqrt{n+1} (n+2)_{k-1} \; \mathcal{B}_{n}(t)^{*} \mathcal{A}_{n+2}(t) \Big \rgroup.
\label{time_dependent_coeff}
\eea
The variance of the quadrature variable is given by
$V_{\theta}= \langle X_{\theta}^{2} \rangle - \langle X_{\theta} \rangle ^2,$
which we will later employ for studying emergence of the squeezed states during the time evolution. 
For later utilization we also quote here the mean photon number $\langle\hat{n}\rangle
 = \hbox{Tr} (a^{\dagger} a\,\rho_{\cal O}(t))$ and its variance $\langle (\Delta\hat{n})^{2}\rangle = 
 \hbox{Tr} (\hat{n}^{2}\,\rho_{\cal O}(t)) - \langle\hat{n}\rangle^{2}$: 
\bea
\braket{\hat{n}} \! \! \! &=&  \! \! \! \mathbf{N}_{1}(t) + \dfrac{x}{4} - \sqrt{x} \; 
\mathrm{Re} \Big \lgroup \mathbf{F}_{1}(t) \Big \rgroup -1 \quad 
\label{avg_photon_number}
\\
\braket{(\Delta \hat{n})^{2}}\!\! &=& \!\!\mathbf{N}_{2}(t) + \dfrac{x}{2} \Big \lgroup  \mathbf{N}_{1}(t) + 
\mathrm{Re} (\mathbf{G}_{2}(t)) - \dfrac{1}{2} \Big \rgroup \! +
 \sqrt{x} \Big \lgroup \mathrm{Re} (\mathbf{F}_{1}(t)) - 2 \; \mathrm{Re} (\mathbf{F}_{2}(t)) \Big \rgroup \nn \\ 
&&- \Big \lgroup \mathbf{N}_{1}(t) - \sqrt{x} \; \mathrm{Re} (\mathbf{F}_{1}(t)) \Big \rgroup^{2}.
\label{variance_num}
\eea

\par

Another dynamical quantity that is useful in the study of `kitten'-like states is the polar phase density  of the 
Husimi $Q$-function [\cite{TMG1993}] obtained via its radial integration on the phase space: 
\beq
\mathcal{Q}(\theta)=\int_{0}^{\infty} Q(\beta, \beta^{*})  \, |\beta|\,\mathrm{d}|\beta|, \quad \beta=|\beta| \exp(i\theta),
\label{QPD}
\eeq
which is a convenient tool for describing the splitting of the $Q$-function. Towards obtaining a series expansion of the phase density
$\mathcal{Q}(\theta)$ we define the  weight factors depending on the polar angle: 
\beq
\mathcal{H}^{(\pm)}_{n,m} (\theta)= \int_{0}^{\infty} H^{(\pm)}_{n,m} (\beta, \beta^{^*}) \, |\beta|\,\mathrm{d}|\beta|
\label{H_theta}
\eeq
that admit, via (\ref{Q_weight}), explicit evaluation as follows:
\bea
\mathcal{H}^{(\pm)}_{n,m} (\theta) &=& \dfrac{1}{\pi \sqrt{n!m!}} 
\sum_{\jmath = 0}^{n}\sum_{\ell=0}^{m} \exp(i (\jmath - \ell) \theta) \exp\Big(-\frac{x \sin^{2} \theta}{4}\Big) 
\binom{n}{\jmath} \binom{m}{\ell} \Big(\frac{x}{4}\Big)^{\frac{n+m-\jmath-\ell}{2}}\nn\\
&&\Big\lgroup P^{(\pm)}_{n+m-\jmath -\ell} \;\Gamma\Big(\frac{\jmath + \ell}{2} + 1\Big)
{}_1F_1 \Big (-\frac{\jmath+\ell+1}{2} ; \frac{1}{2}; -\frac{x \cos^{2} \theta}{4}\Big)\nn\\
&&-P^{(\mp)}_{n+m-\jmath -\ell} \;\sqrt{x}\,\cos \theta\;\Gamma\Big(\frac{\jmath + \ell +3}{2}\Big)
{}_1F_1 \Big (-\frac{\jmath+\ell}{2}; \frac{3}{2}; -\frac{x \cos^{2} \theta}{4}\Big)\Big\rgroup,
\label{H_nm}
\eea
where we have used the integration given by
\bea
\int_{0}^{\infty} r^p \exp(-(r \pm \mathcal{X})^2) dr &=&  \frac{1}{2}\, 
\Gamma \Big (\frac{p+1}{2} \Big) {}_1F_1 \Big (-\frac{p}{2} ; \frac{1}{2}; - \mathcal{X}^{2}\Big)  \nn \\ 
&& \mp \;\mathcal{X}\;
\Gamma \Big(\frac{p}{2} + 1\Big) 
{}_1F_1 \Big (\frac{1-p}{2} ; \frac{3}{2}; -\mathcal{X}^{2}\Big).
\label{r_shift_int}
\eea
The above compendium of expressions (\ref{QPD}, \ref{Q_function}, \ref{H_nm}) now provides the evolution of 
the polar phase density  $\mathcal{Q}(\theta)$ for our initial state (\ref{t0quasiBell})
\bea
\mathcal{Q}(\theta) &=& \dfrac{1}{2}  |\mathcal{C}_{0}(t)|^{2} \mathcal{H}^{(+)}_{0,0}(\theta)
 + \mathrm{Re} \Big \lgroup \mathcal{C}_{0}(t)^{*} 
\sum_{n=1}^{\infty} \Big( \mathcal{A}_{n}(t) \mathcal{H}^{(-)}_{0,n-1}(\theta)\nn  \\
&&+ \mathcal{B}_{n}(t) \mathcal{H}^{(+)}_{0,n}(\theta) \Big) + 
\sum_{n,m=1}^{\infty} \! \mathcal{A}_{n}(t)^{*} 
\mathcal{B}_{m}(t) \mathcal{H}^{(-)}_{n-1,m}(\theta) \Big \rgroup \nn \\
&& + \dfrac{1}{2}\! \sum_{n,m=1}^{\infty} \!\Big( \mathcal{A}_{n}(t)^{*} 
\mathcal{A}_{m}(t) \mathcal{H}^{(+)}_{n-1,m-1}(\theta) 
+\mathcal{B}_{n}(t)^{*} \mathcal{B}_{m}(t) \mathcal{H}^{(+)}_{n,m}(\theta) \Big).
\label{Q_PD}
\eea
\section{Delocalization on the phase space}
\label{DelocalQuantum}
\setcounter{equation}{0}
\subsection{Wehrl entropy}
\label{Wehrl}
\begin{figure}[H]
\captionsetup[subfigure]{labelformat=empty}
\subfloat[(a)]{\includegraphics[width=4cm,height=3cm]{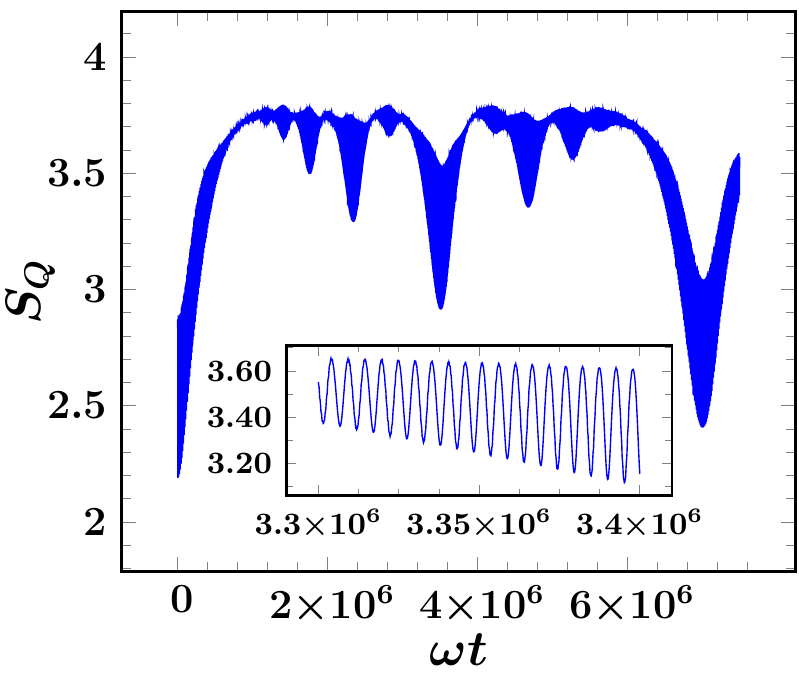}} 
\captionsetup[subfigure]{labelformat=empty}
\subfloat[(b)]{\includegraphics[width=4cm,height=3cm]{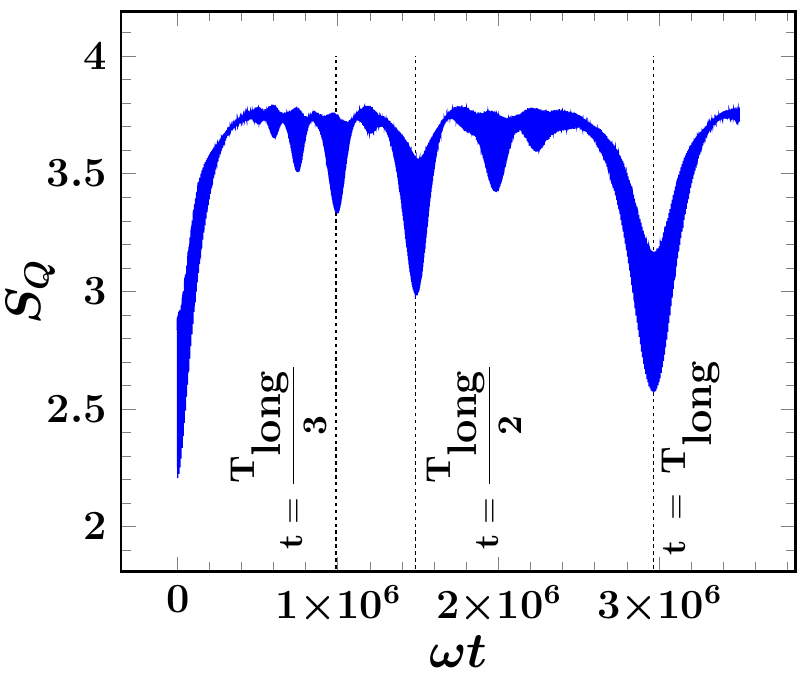}} 
\captionsetup[subfigure]{labelformat=empty}
\subfloat[(c)]{\includegraphics[width=4cm,height=3cm]{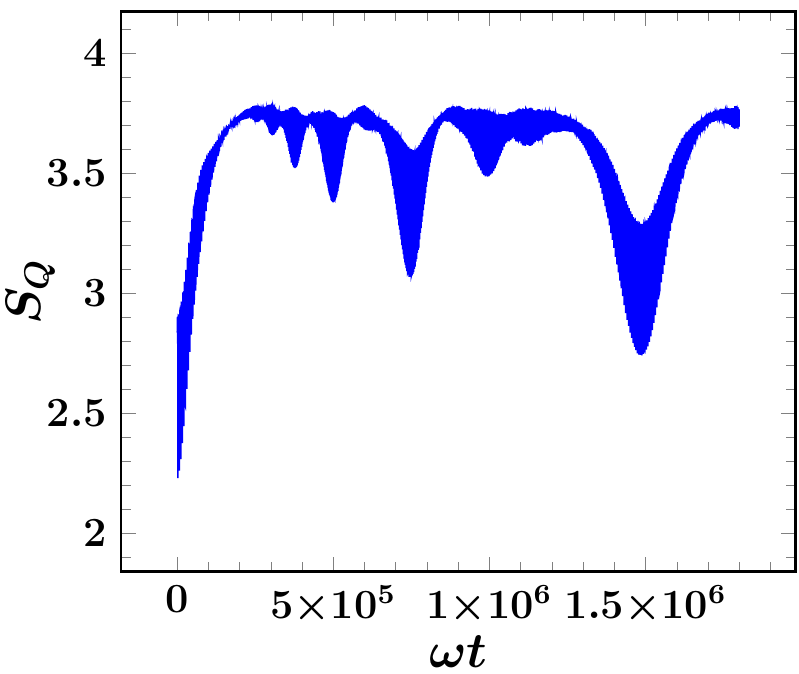}}
\captionsetup[subfigure]{labelformat=empty}
\subfloat[(d)]{\includegraphics[width=4cm,height=3cm]{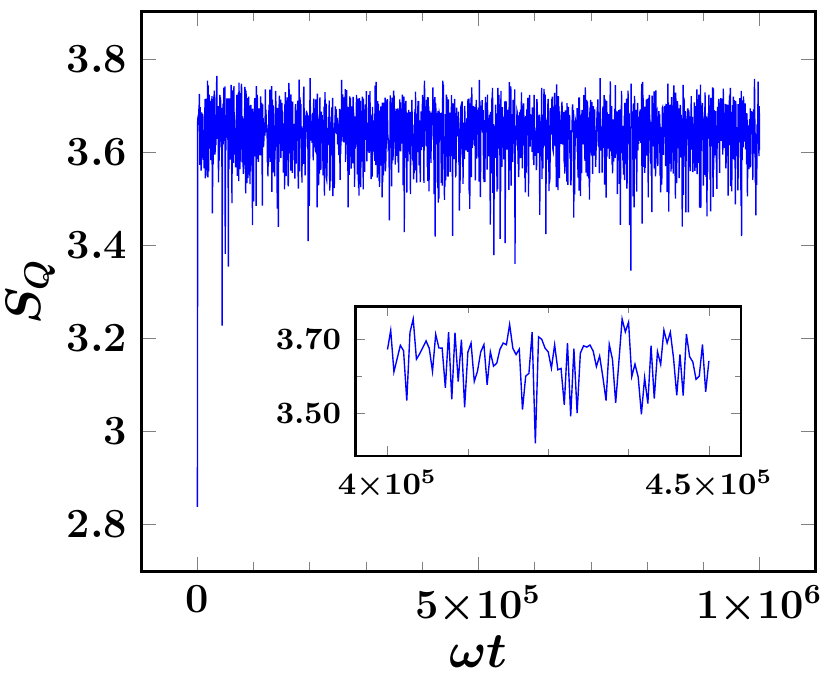}} 
\caption{The time evolution of the Wehrl entropy $S_{Q}$ for the parameters $\Delta=0.8\, \omega,\alpha=2.5$, and various values of 
$\lambda$: (a) $0.008\, \omega$, (b) $0.01\, \omega$, (c) $0.012\, \omega$, (d) $0.1\, \omega$. In the cases (a)-(c) a quasi periodic behavior is observed,
whereas in (d) the randomized phase relationship between a large number of modes leads to a stochastic stabilization of $S_{Q}$.}
\label{Wehrl_Time}
\end{figure}
\begin{figure}[H]
\begin{minipage}[c]{0.35\textwidth}
    \includegraphics[width=6cm,height=4.5cm]{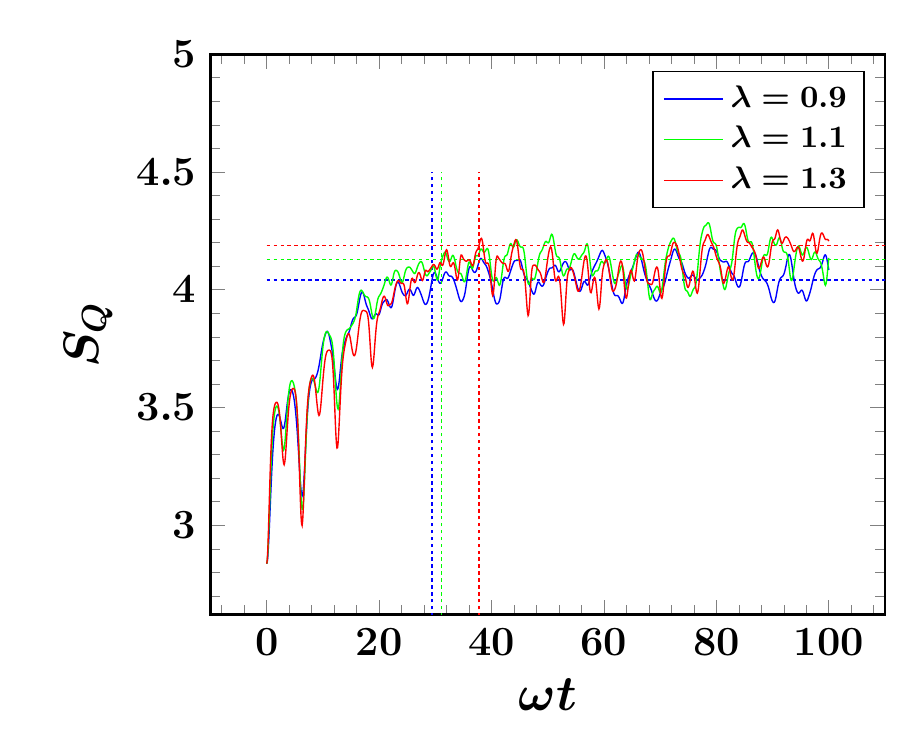}
  \end{minipage}\hfill
  \begin{minipage}[c]{0.6\textwidth}
\caption{The  initial rise of the Wehrl entropy $S_{Q}$ for the parameters $\Delta=0.5\, \omega , \alpha =2.5$, and the coupling strength $\lambda = 0.9 \,\omega\hbox{(blue)}, 1.1\, \omega\hbox{(green)}, 1.3 \, \omega \hbox{(red)}$. The horizontal dotted lines (for the respective colors) measure the corresponding 
time-averaged values of $S_{Q}$ that read $4.0420, 4.1280, 4.1890$ in the said order. The averaging is done for a time interval $\omega t=4000$ in each case.
The vertical dotted lines signify the first crossing of the averaged $S_{Q}$ by the corresponding time evolution graph. This is regarded as the measure of
$  T_{\mathrm{ent. prod.}}$. The arbitrariness in this determination can be improved by a local smoothing operation.
For the respective coupling strengths these estimates read $29.40, 31.11, 37.76$ in the said order.}
\label{Wehrl_initial rise}
\end{minipage}
\end{figure}

% % % % % % % % % %
The Wehrl entropy defined as [\cite{W1978}]
\beq
S_{Q}=-\int Q(\beta, \beta^{*}) \,\log Q(\beta, \beta^{*}) \, \mathrm{d}^{2}\beta
\label{WehrlDef}
\eeq  
is an information-theoretic measure estimating the delocalization of the system in the  oscillator
phase space. It is considered [\cite{BKK1995}] as a count of an equivalent number of widely
separated coherent states necessary for covering the existing phase space occupation of 
the coupled oscillator. Being subject to the restriction originating from the Heisenberg uncertainty principle, 
the Wehrl entropy (\ref{WehrlDef}) is a positive definite quantity [\cite{W1978}]. 
In the present case we employ the definition (\ref{WehrlDef}) and  the evolution  (\ref{Q_function}) 
of the $Q$-function  to numerically study  the long-range time dependence of $S_{Q}$
 for various values of the coupling strength. 
 We note that here and hereafter the  time is measured in the natural unit: $\omega^{-1}$. 
 We observe the following properties: ({\sf{i}})
In the  long time limit $t \gtrsim (x^{2} \widetilde{\Delta})^{-1}$ the quasi periodicity of the Wehrl entropy 
is manifest in the coupling strength regime $\lambda \lesssim 0.05 \,\omega$, where the Laguerre polynomials $L_{n}^{(j)}(x)$
are well-approximated by their quadratic components $O(x^{2})$. Frequency modes $O(x^{2} \widetilde{\Delta})$ 
and their harmonics  produced via the interaction now give rise to the quasi periodicity of  
$S_{Q}$, where the long range time period maintains the property: $T_{\mathrm{long}} \propto  
\lambda^{-4}\, \exp (x/2)$ (Figs. \ref{Wehrl_Time} (a)-(c)). For a smaller value of the qubit frequency 
$\Delta$ the quasi periodic behavior persists for a comparatively higher coupling strength $\lambda$. This  
 follows from the requirement that for the quasi periodicity to hold, the phase change caused by the higher order fluctuations 
  $\{O(x^{n})| n > 2\}$ during the time span $T_{\mathrm{long}}$ is to be negligibly 
small: $T_{\mathrm{long}}\, x^{3} \widetilde{\Delta} \ll 1$. The time period observed in 
Figs. \ref{Wehrl_Time}(a)-(c) are noted below:  $T_{\mathrm{long}} = 6.9914 \times 10^{6}$ (for $\lambda = 0.008\,\omega$), 
 $T_{\mathrm{long}} = 2.9568 \times 10^{6}$ (for $\lambda = 0.010\, \omega$), $T_{\mathrm{long}} = 
 1.4743 \times 10^{6}$ (for $\lambda = 0.012\, \omega$), respectively. The near equality of the 
product $T_{\mathrm{long}}\, \lambda^{4}\, \exp (-x/2)$ in the respective cases ($0.02863, 0.02956, 0.03056)$ 
validates our argument that the quantum fluctuations $O(x^{2})$ produce the observed long range time period.
In the instance of nonlinear Kerr-like medium similar behavior in the time evolution of  $S_{Q}$ was previously 
noticed [\cite{JO1994}], where its local minima corresponded with the formations of finite superposition of 
coherent states. In the present bipartite interacting model, however, the qubit-oscillator interaction 
superimposes short time span fluctuations of the frequency $O(x \widetilde{\Delta})$ on the long range  
oscillations that now act as an envelope of the total time evolution of the Wehrl entropy. This introduces important distinctions to the present  model. We will  discuss this in the 
Subsec. \ref{kitten}. ({\sf{ii}}) In the ultra-strong coupling regime $\lambda \gtrsim  0.1\, \omega$ all 
frequency modes $\{O(x^{n} \widetilde{\Delta})| n= 0, 1, \ldots\}$ and their harmonics arise. Random phase differences 
between  a large number of incommensurate modes cause the resultant interference to average out, while ensuring an 
effective stabilization of the occupation of the phase space after an 
initial build up (Fig. \ref{Wehrl_Time} (d)). The rapid high frequency $O(\omega)$ fluctuations are of small amplitude:
$|\Delta S_{Q}|/ S_{Q} \ll 1$, and may be removed by a suitable coarse graining process [\cite{JJ2007}].
In this regime it is  observed that for a fixed initial state parameter $\alpha$ the time-averaged value of 
the Wehrl entropy gradually increases with increasing coupling strength. Following (\ref{avg_photon_number}) it is 
evident that   higher coupling strength leads to an enhancement of the photon expectation value $\braket{\hat{n}}$ that causes 
a wider spread of the $Q$-function resulting in an increment in $S_{Q}$.   
\begin{figure}[]
\begin{center}
 \captionsetup[subfigure]{labelformat=empty}
\subfloat[(a)]{\includegraphics[width=3.3cm,height=3.3cm]{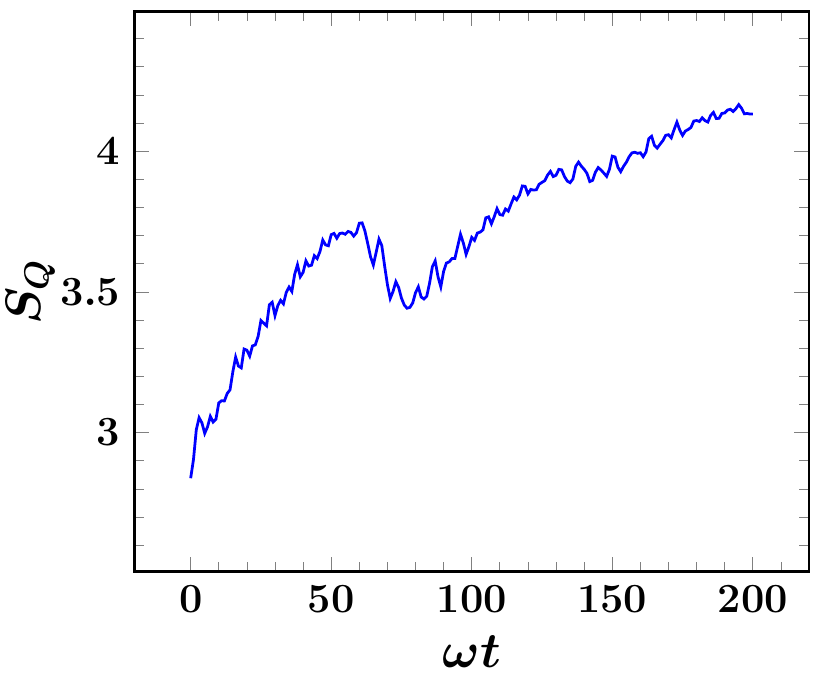}} \quad
\captionsetup[subfigure]{labelformat=empty}
\subfloat[(b)]{\includegraphics[width=3.3cm,height=3.3cm]{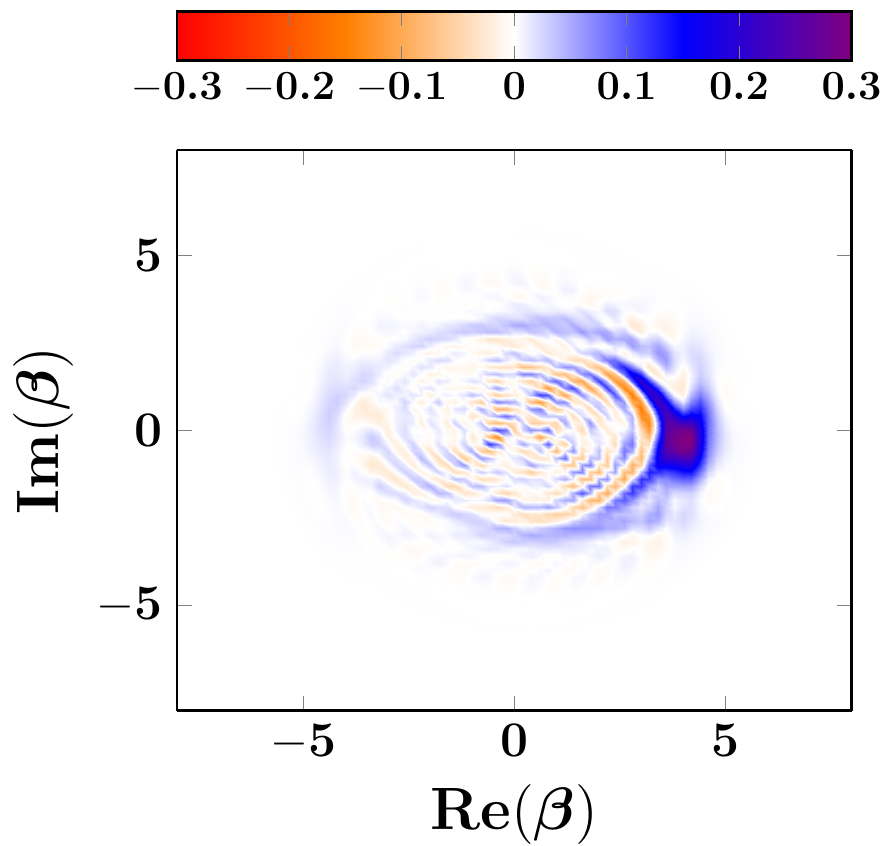}} \quad
\captionsetup[subfigure]{labelformat=empty}
\subfloat[(c)]{\includegraphics[width=3.3cm,height=3.3cm]{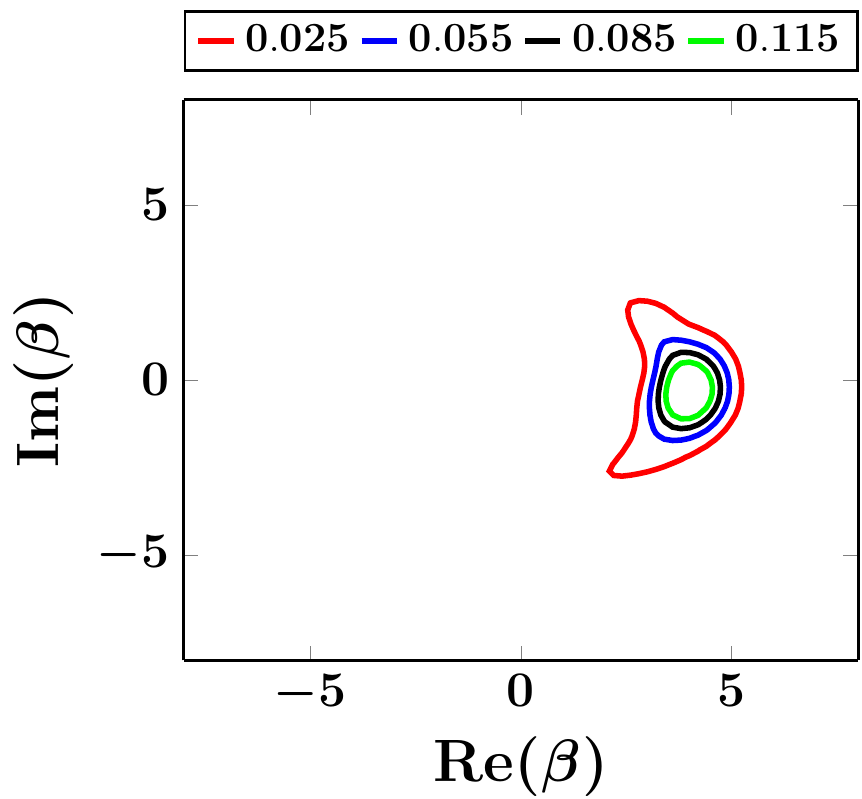}} \quad
\captionsetup[subfigure]{labelformat=empty}
\subfloat[(d)]{\includegraphics[scale=0.4]{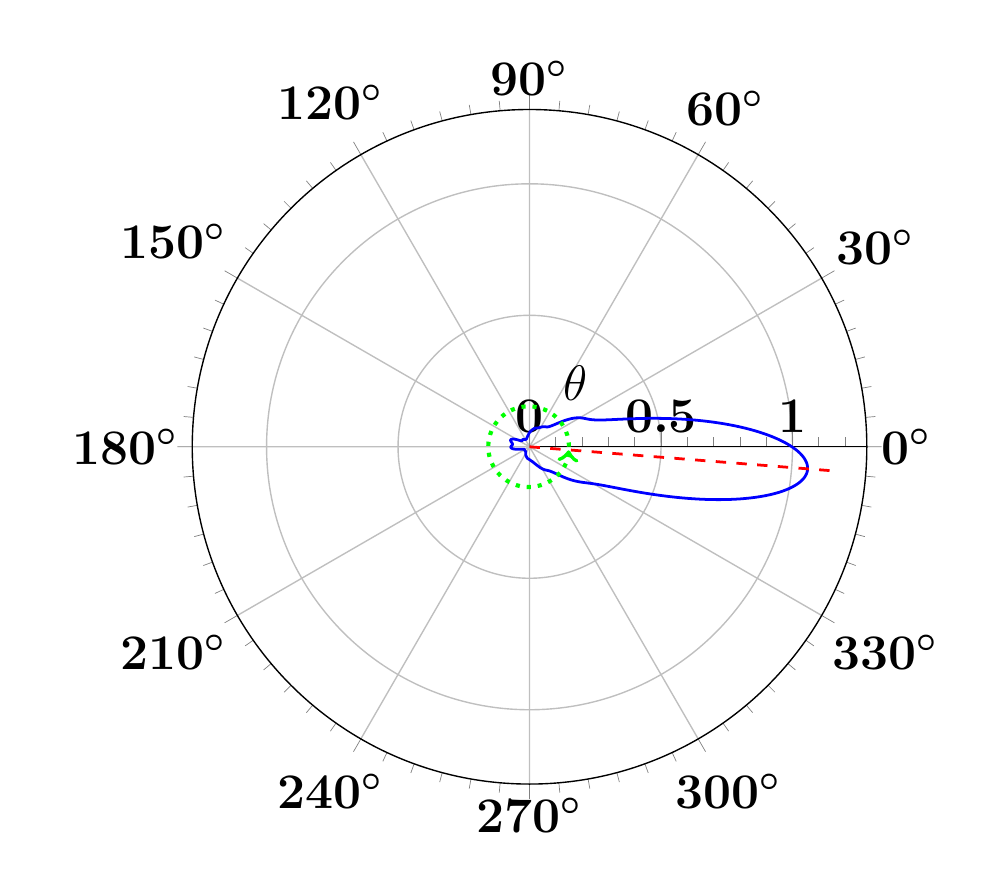}} 
\caption{For the parametric values of $\Delta = \omega , \lambda=0.3\, \omega, \alpha =4 $ (a) refers to the time evolution of 
$S_{Q}$. At its local minimum ($\omega t=77$) the Wigner $W$-distribution (b), the Husimi $Q$-function (c), and the polar plot 
(w.r.t. $\theta$ ) of the phase density $\mathcal{Q}(\theta)$ (d) are given, respectively. The localized single peak of the phase density 
$\mathcal{Q}(\theta)$ occurs at $\theta=355.49^{\circ}$.}
 \label{single_peak}
 \end{center}	
\end{figure} 
({\sf{iii}}) Sufficiently localized states are transiently observed at large macroscopic values of the coherent state amplitude 
$\alpha$  even in the ultra-strong coupling domain. These states are realized (Fig.\ref{single_peak}) at the local minimum of the Wehrl entropy $S_{Q}$. Physically, the energy exchange between the qubit and the oscillator degrees of freedom induces the revival and collapse of the qubit density matrix elements. A revival of the qubit matrix element indicates that the oscillator has, reciprocally, less energy  available to it. This acts as a constraint on its delocalization on the phase space.  Moreover, a large value of the amplitude $\alpha$ signifies comparatively higher magnitude of energy residing  in the interference pattern (Fig.\ref{single_peak} (b)) of the quantum oscillations. Qualitatively, when the energy associated with the stochastic randomization of the modes is less than the energy of the coherent interference pattern, a localization on the phase space takes place 
and, consequently, a relative decrease in $S_{Q}$ develops. The quantum interference pattern (Fig.\ref{single_peak} (b)) has the shape of a localized peak with `twisted arms', where oscillations develop perpendicular to these arms causing a transport of energy necessary for the localization. The phase density diagram (Fig. \ref{single_peak} (d)) shows that the initial state  (\ref{t0quasiBell}) consisting of two almost maximally mixed Gaussian peaks coalesce at the local minimum of $S_{Q}$ to produce  a partially pure transient  state of the oscillator with its von Neumann entropy given by $S = 0.5883$.
 ({\sf{iv}}) The fast initial rise of $S_{Q}$
 in the ultra-strong coupling domain $\lambda \sim \omega$ may be understood as follows. Increased 
qubit-oscillator coupling leads to the generation of all high-frequency quantum fluctuation modes. The phase randomization 
of the modes of incommensurate frequencies results in the rapid initial spreading on the phase space, and consequent fast  
production of $S_{Q}$. Once the modes are statistically populated, the Wehrl entropy $S_{Q}$, except for high frequency 
$O(\omega)$ quantum fluctuations of relatively small amplitude, 
maintains an almost stationary value. The initial production time of the Wehrl entropy $T_{\hbox{\tiny{ent. prod.}}}$
follows from the asymptotic behavior of associated Laguerre polynomials $L_n^{(j)}(x)$ at large 
$n \gg 1$, fixed $j$, and $x > 0$ [\cite{S1975}]:
\beq
L_n^{(j)}(x) = \frac{n^{\frac{j}{2}-\frac{1}{4}}}{\sqrt{\pi}} 
\frac{\exp(x/2)}{x^{\frac{j}{2} + \frac{1}{4}}} \cos\left(2 \sqrt{nx}- 
\frac{\pi}{2} \left(j +\frac{1}{2}\right) \right) + O\left(n^{\frac{j}{2}-\frac{3}{4}}\right).
\label{Lagurre_asymptotic}
\eeq
The asymptotic limit of the energy eigenvalues (\ref{Hn_eigenstate}) is now readily obtained:
\beq
\mathcal{E}^{(\pm)}_{n (\gg 1)} = \omega\left(n \pm 1 -\frac{1}{2} - \frac{x}{4}\right) \mp \frac{\Delta}{\sqrt{\pi}}\,
(n x)^{-\frac{1}{4}} \cos \left(2 \sqrt{nx}- \frac{\pi}{4}\right)+ O(n^{-\frac{1}{2}}),
\label{E_asymptotic}
\eeq
where the leading interaction-generated part $O\Big(n^{-\frac{1}{4}}\Big)$ on the rhs provides the 
effective statistical stabilization of the occupation in the phase space. Recognizing this, the coupling 
strength dependence of the typical time scale  for the generation of $S_{Q}$ is now given by 
 $T_{\mathrm{ent. prod.}}\, \frac{\Delta}{\sqrt{\pi}}\,(n x)^{-\frac{1}{4}} \sim 1 \Rightarrow 
 T_{\mathrm{ent. prod.}} \propto \sqrt{\lambda}$. The rapid initial  increase of the Wehrl entropy is described in Fig. 
 \ref{Wehrl_initial rise}.  For the data presented in  Fig.  \ref{Wehrl_initial rise} the proportionality constant 
 $T_{\mathrm{ent. prod.}} /\sqrt{\lambda}$ read $30.99, 29.66, 33.11$  for the coupling strengths 
 $\lambda=0.9\, \omega, 1.1\, \omega, 1.3\, \omega$, respectively. The 
 discrepancy ($\sim 6\%$) in the observed data occurs since the local fluctuations in the time evolution of $S_{Q}$ play an 
 important role in determining $T_{\mathrm{ent. prod.}}$. A suitable coarse-graining process [\cite{JJ2007}] to smooth 
 the high frequency fluctuations may be adopted for fuller agreement.
 
\subsection{Kitten states and multiple time scales}
\label{kitten}
For the Kerr-type nonlinear self-interacting photonic models, the local minima in the time evolution of $S_{Q}$ 
 of an initial coherent state are associated [\cite{JO1994}] with transient formations of the superposition of 
a finite number of coherent states [\cite{MTK1990}-\cite{MBWI2001}] maintaining a uniform angular separation on the
complex plane. These superpositions are realized at rational submultiples of the time period of  $S_{Q}$.  
Recently such nonclassical superposition of multiple coherent states in a Kerr medium has been experimentally achieved 
[\cite{Kirchmair2013}]. Formation of cat-like states in a finite dimensional bosonic system that admits applying a 
displacement operator on its ground state has also been studied [\cite{MPPN2014}] in a Kerr medium.

\par

\begin{figure}[H]
\captionsetup[subfigure]{labelformat=empty}
\subfloat[(a$_{1}$)]{\includegraphics[width=3cm,height=3cm]{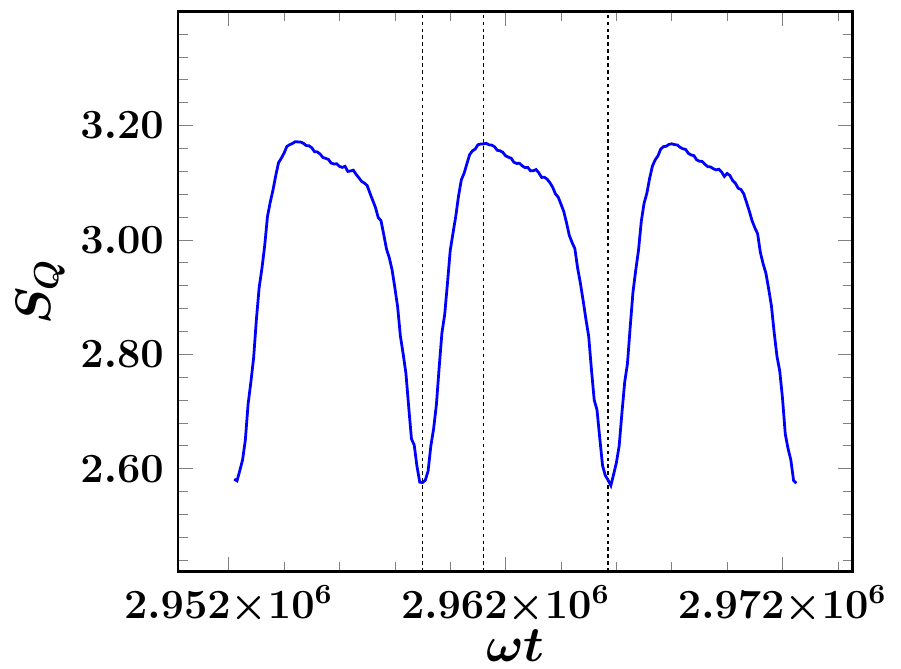}} 
\captionsetup[subfigure]{labelformat=empty}
\subfloat[(a$_{2}$)]{\includegraphics[width=3cm,height=3cm]{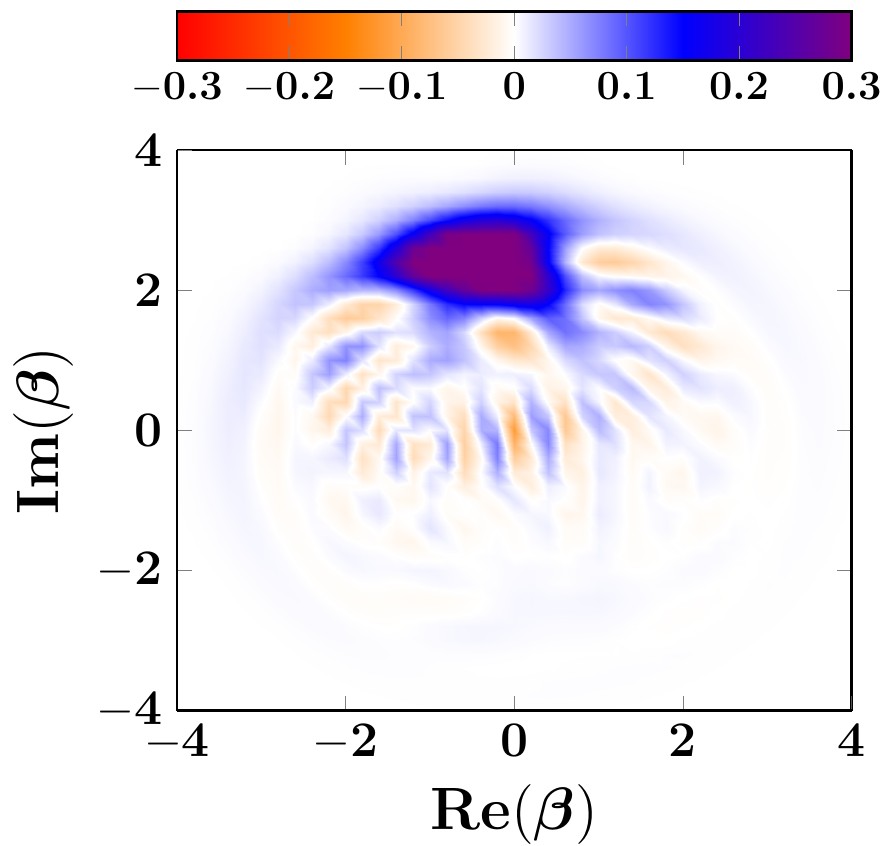}} 
\captionsetup[subfigure]{labelformat=empty}
\subfloat[(a$_{3}$)]{\includegraphics[width=3cm,height=3cm]{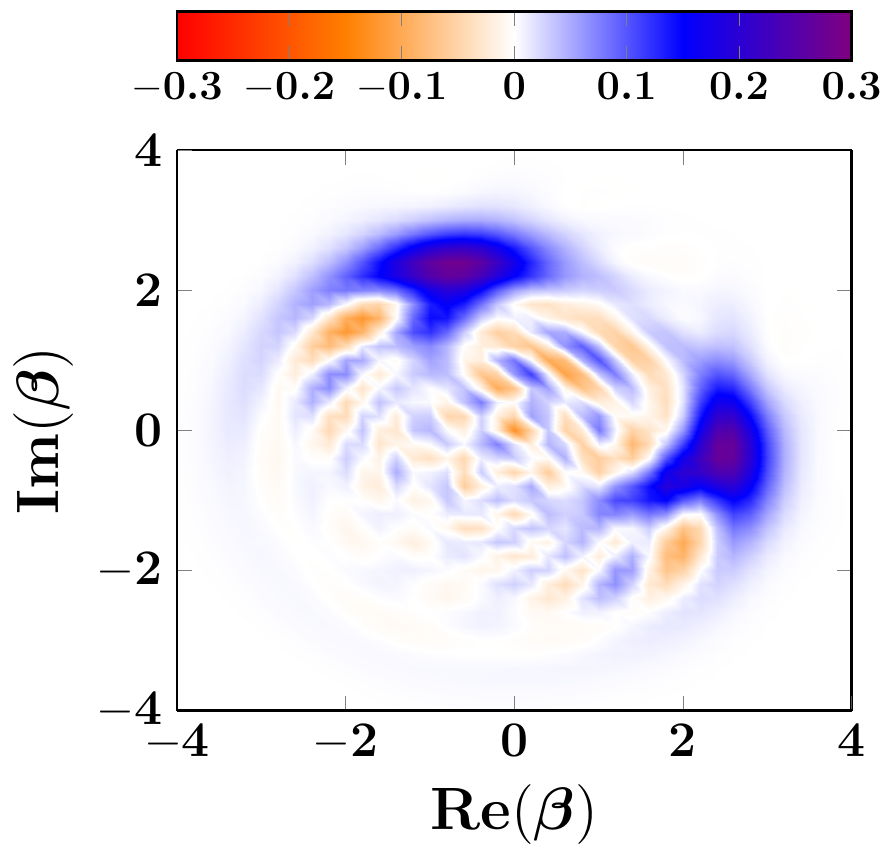}}
\captionsetup[subfigure]{labelformat=empty}
\subfloat[(a$_{4}$)]{\includegraphics[width=3cm,height=3cm]{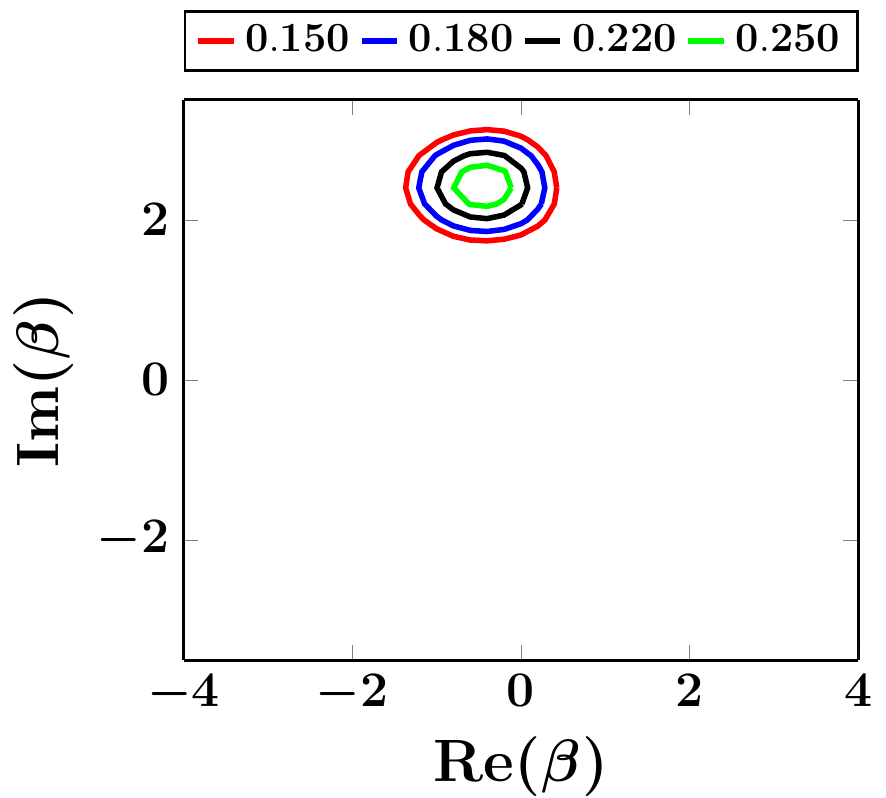}} 
\captionsetup[subfigure]{labelformat=empty}
\subfloat[(a$_{5}$)]{\includegraphics[width=3cm,height=3cm]{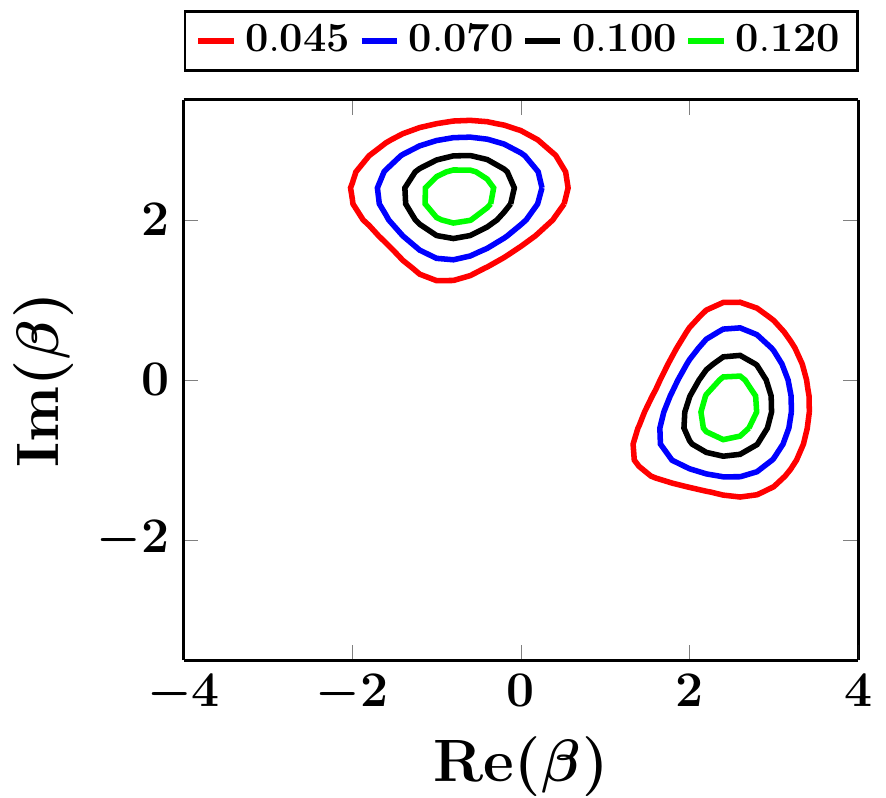}} \\
\captionsetup[subfigure]{labelformat=empty}
\subfloat[(b$_{1}$)]{\includegraphics[width=3cm,height=3cm]{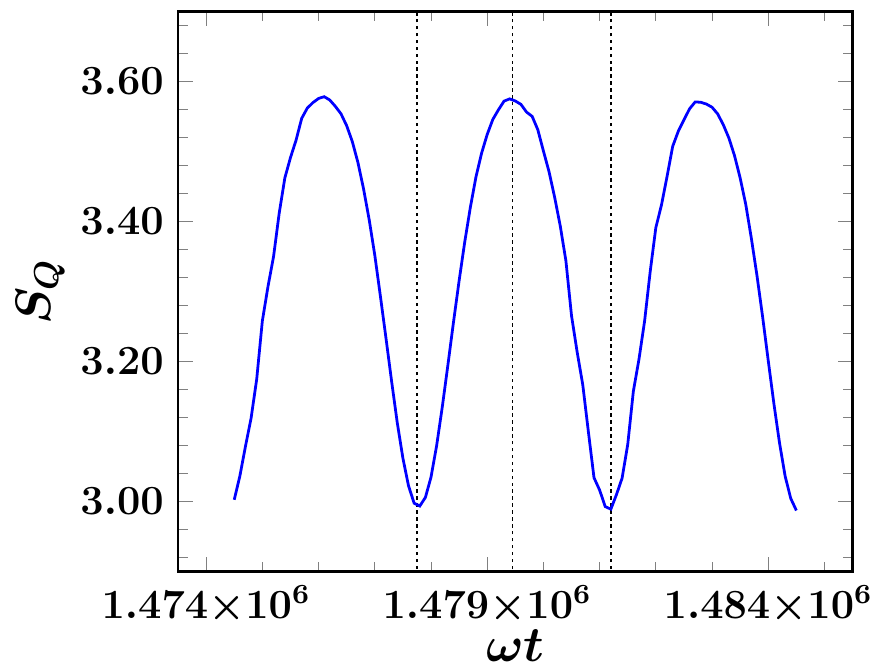}}
\captionsetup[subfigure]{labelformat=empty}
\subfloat[(b$_{2}$)]{\includegraphics[width=3cm,height=3cm]{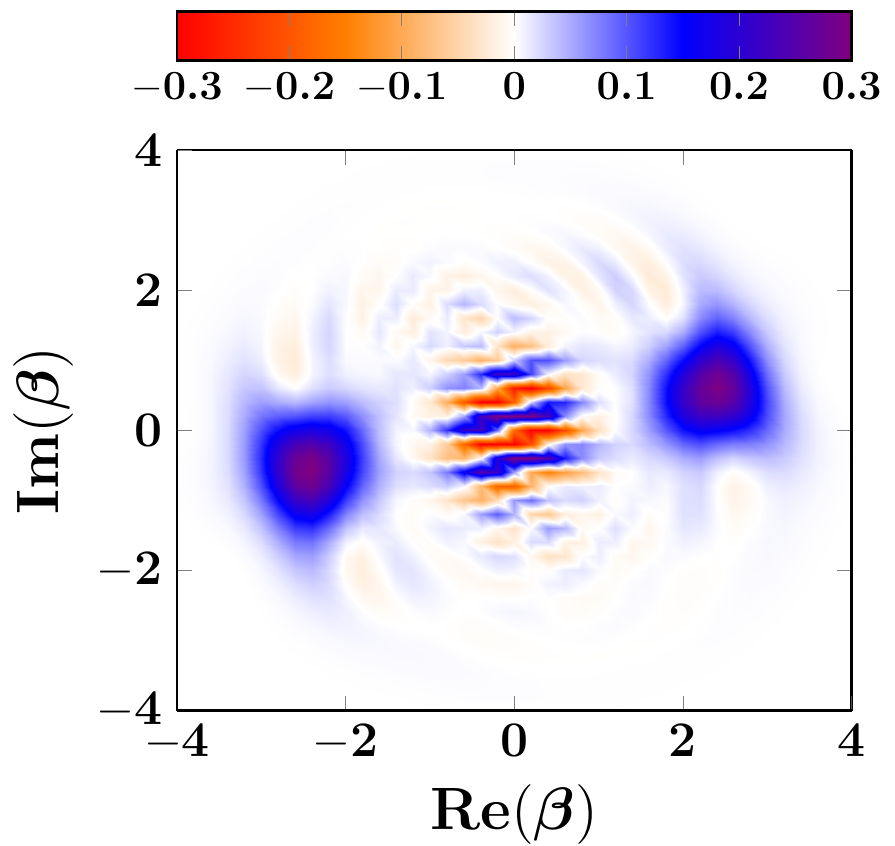}} 
\captionsetup[subfigure]{labelformat=empty}
\subfloat[(b$_{3}$)]{\includegraphics[width=3cm,height=3cm]{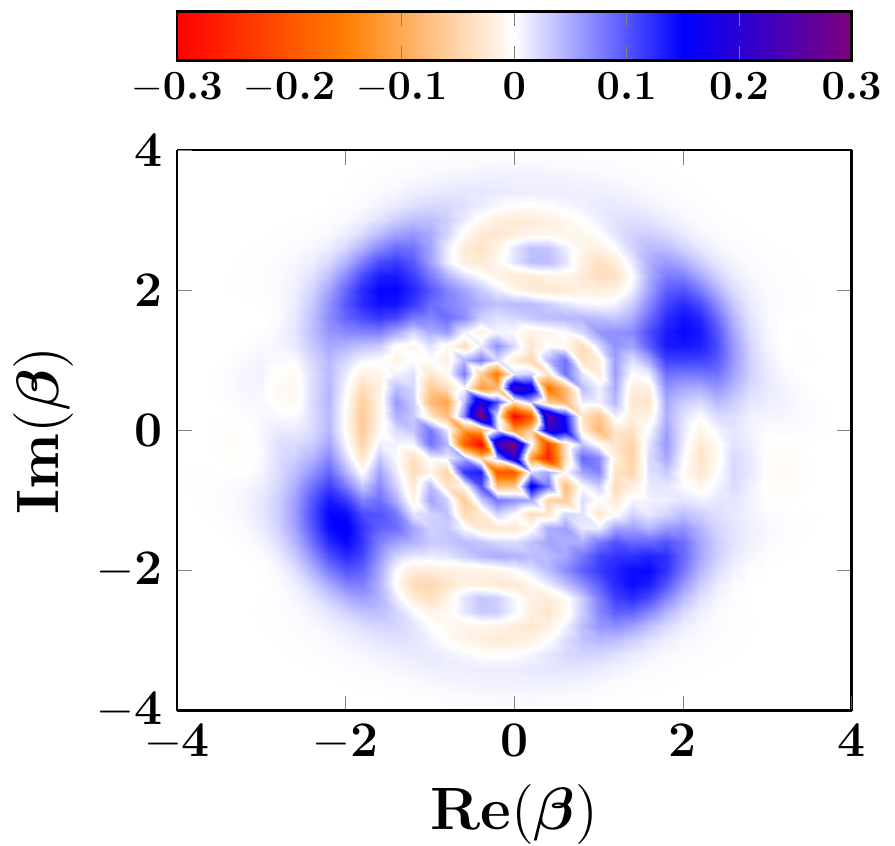}} 
\captionsetup[subfigure]{labelformat=empty}
\subfloat[(b$_{4}$)]{\includegraphics[width=3cm,height=3cm]{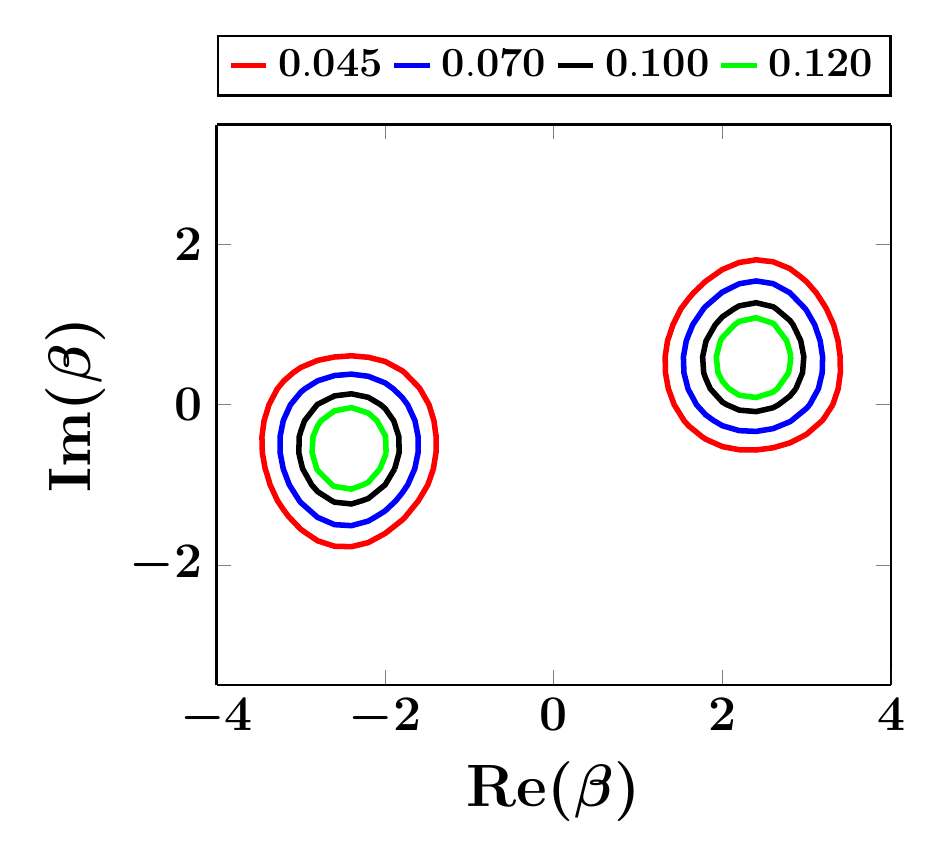}}
\captionsetup[subfigure]{labelformat=empty}
\subfloat[(b$_{5}$)]{\includegraphics[width=3cm,height=3cm]{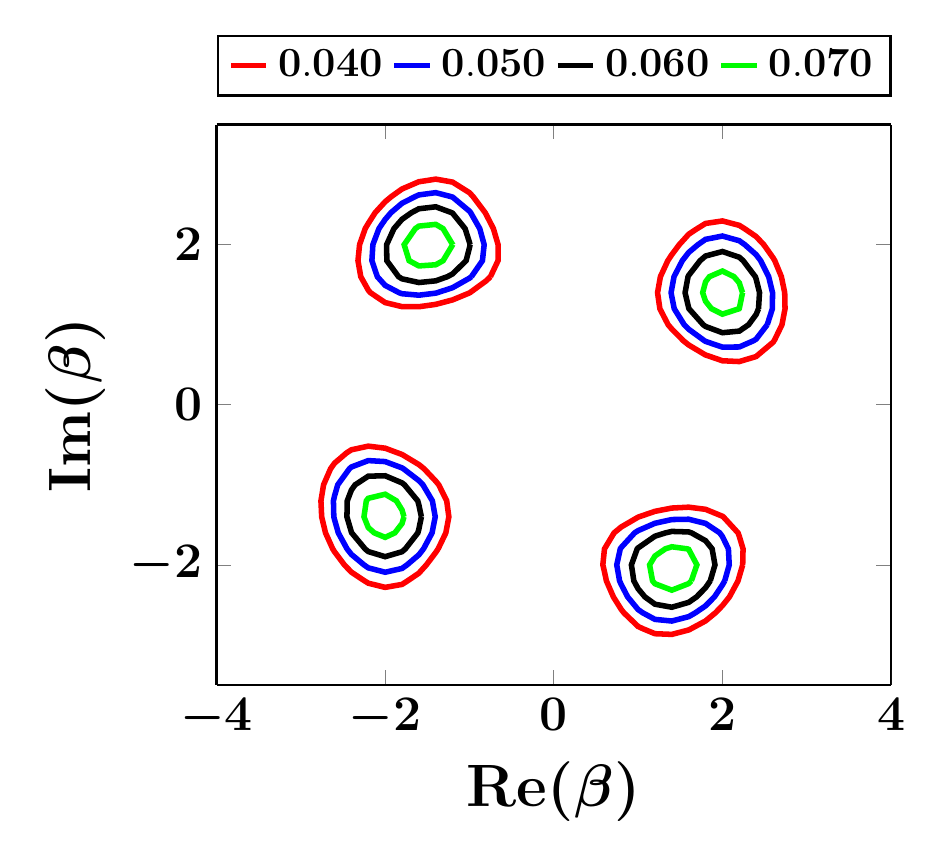}} \\
\captionsetup[subfigure]{labelformat=empty}
\subfloat[(c$_{1}$)]{\includegraphics[width=3cm,height=3cm]{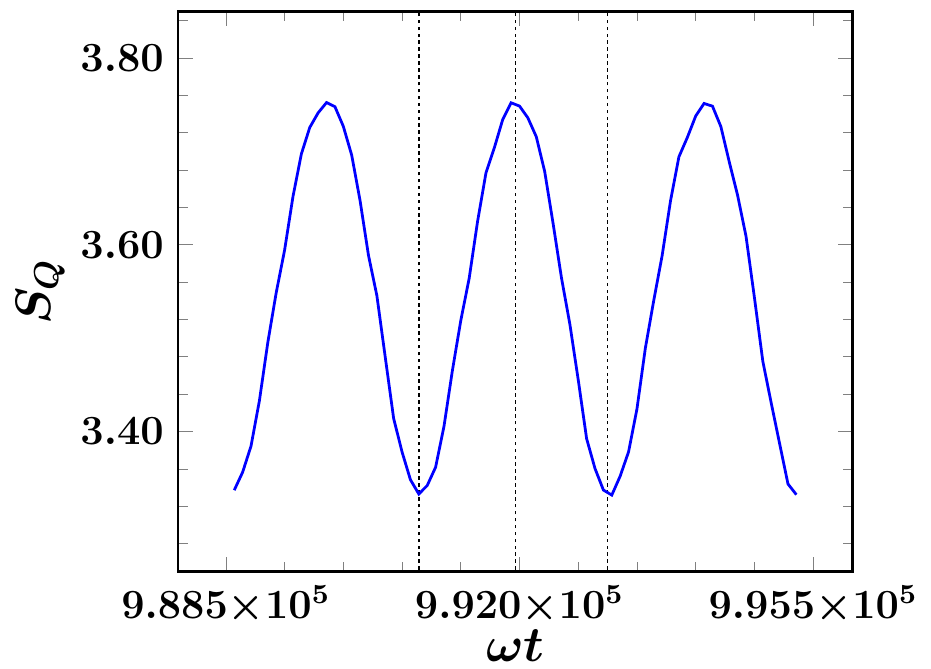}}
\captionsetup[subfigure]{labelformat=empty}
\subfloat[(c$_{2}$)]{\includegraphics[width=3cm,height=3cm]{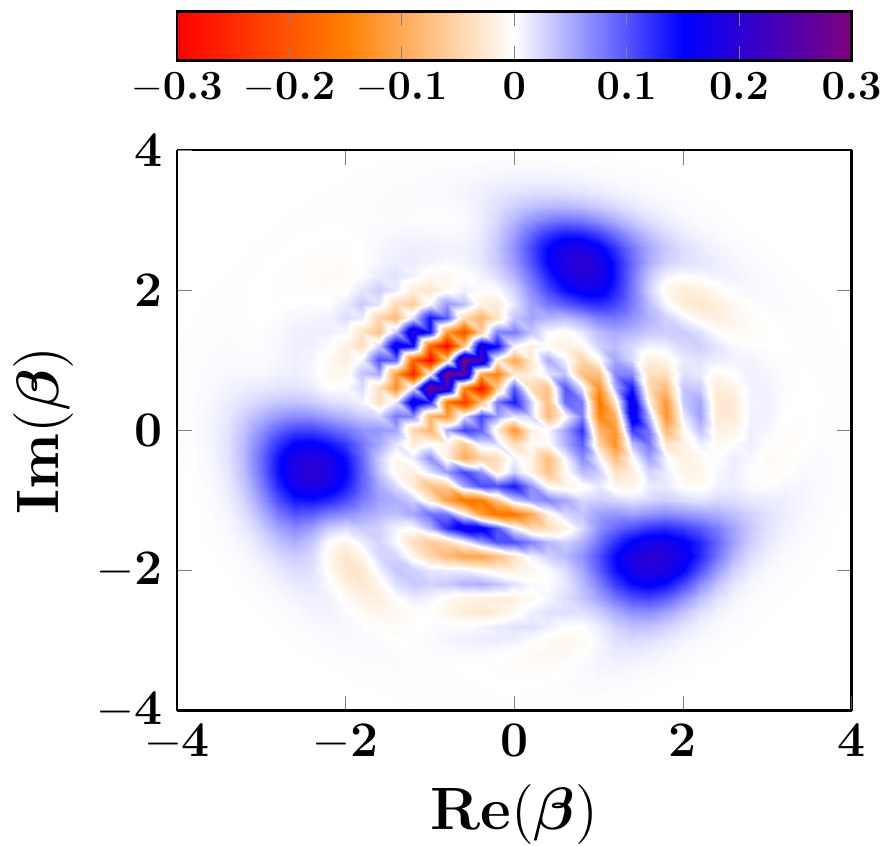}} 
\captionsetup[subfigure]{labelformat=empty}
\subfloat[(c$_{3}$)]{\includegraphics[width=3cm,height=3cm]{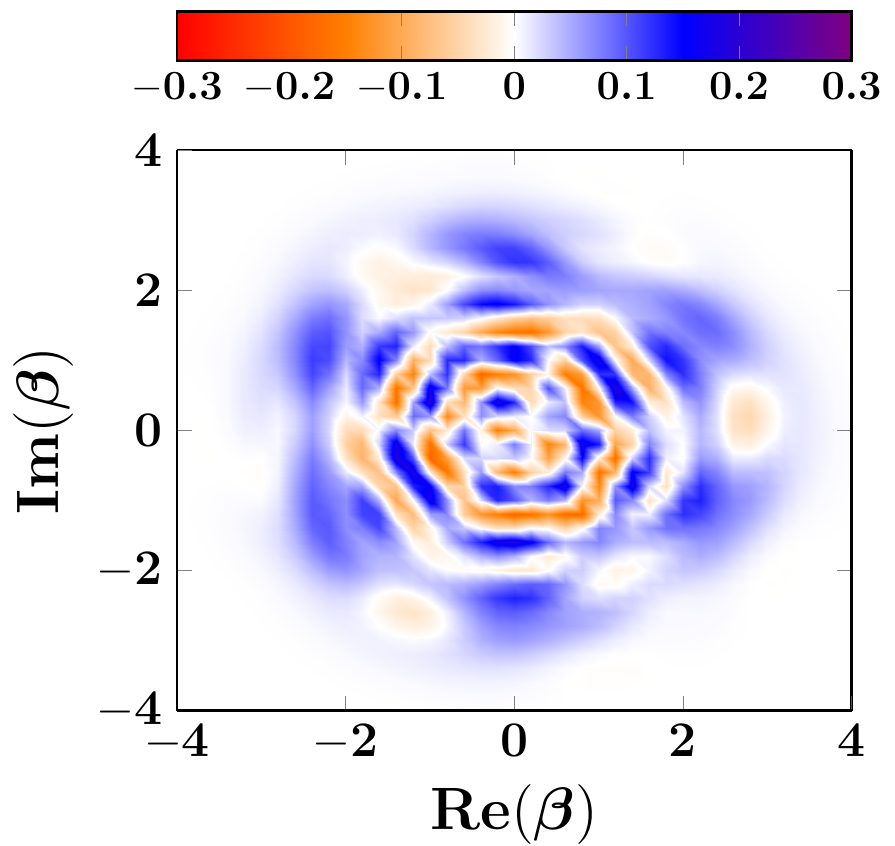}} 
\captionsetup[subfigure]{labelformat=empty}
\subfloat[(c$_{4}$)]{\includegraphics[width=3cm,height=3cm]{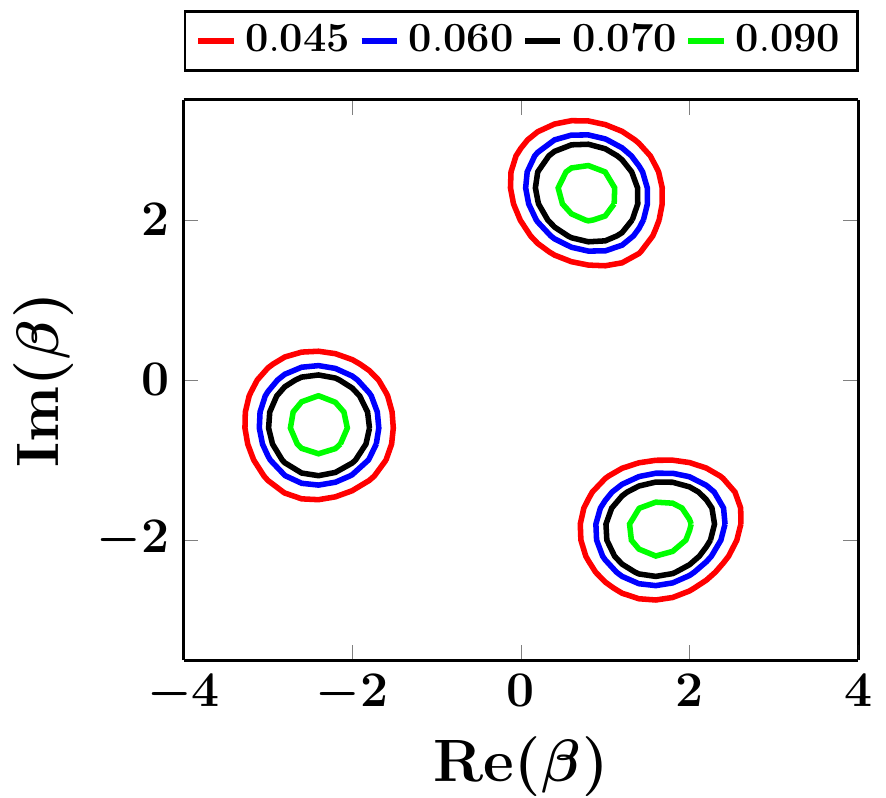}}
\captionsetup[subfigure]{labelformat=empty}
\subfloat[(c$_{5}$)]{\includegraphics[width=3cm,height=3cm]{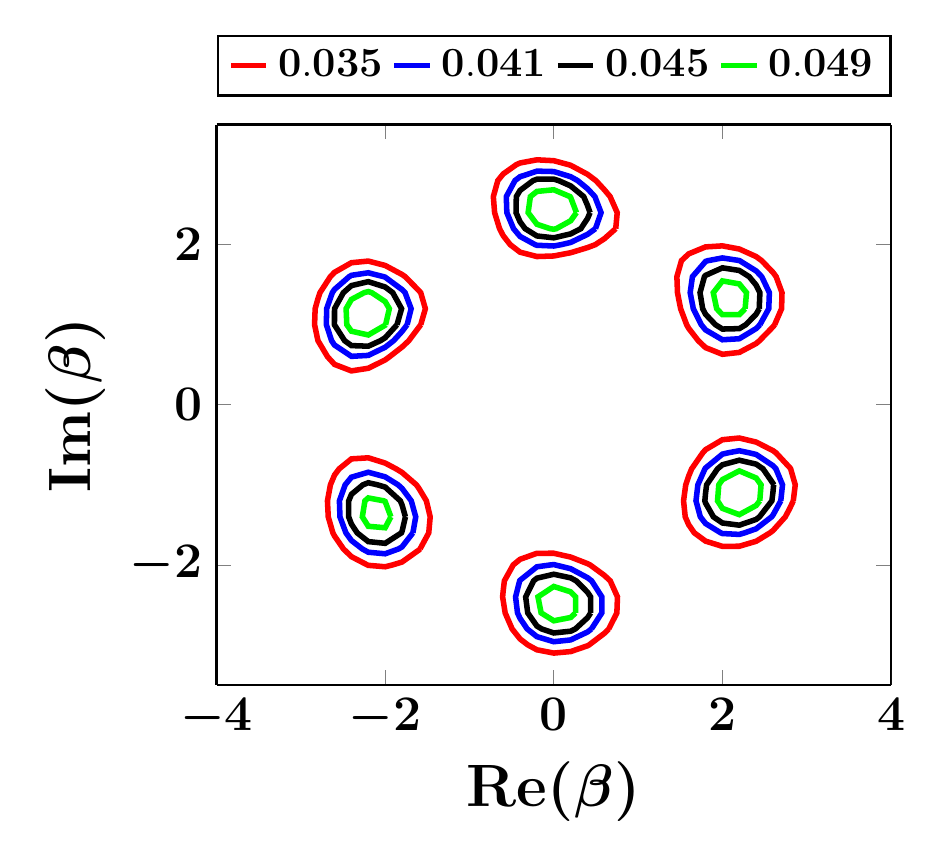}}
\caption{Transient formations of `kitten' states are observed for the parameters $\Delta=0.8\, \omega, 
\lambda=0.01 \, \omega,\alpha=2.5$.  The rows $(\{\hbox{a}\}, \{\hbox{b}\}, \{\hbox{c}\})$ describe the results 
at times $T_{\mathrm{long}},T_{\mathrm{long}}/2, T_{\mathrm{long}}/3$, respectively. The column 
(a$_{1}$, b$_{1}$, c$_{1}$) marks the short time period oscillations (frequency $O(x \widetilde{\Delta})$) of the 
Wehrl entropy $S_{Q}$. The respective times corresponding to the local minima in (a$_{1}$, b$_{1}$, c$_{1}$) are 
$2.9590 \times 10^{6}, 1.4778 \times 10^{6}, 9.9080 \times 10^{5}$, whereas the times of the subsequent local maxima, in turn, read
$2.9612 \times 10^{6}, 1.4794 \times 10^{6}, 9.9195 \times 10^{5}$. Between the minima and the maxima of the short time 
period oscillations, a doubling of the number of `kittens' 
is observed. The columns (a$_{2}$, b$_{2}$, c$_{2}$) and (a$_{3}$, b$_{3}$, c$_{3}$) specify the Wigner 
$W$-distribution at the minima and the maxima of the said short time period oscillation, respectively. 
Corresponding results for the smoothed $Q$-function are subsequently given in the columns (a$_{4}$, b$_{4}$, c$_{4}$) and 
(a$_{5}$, b$_{5}$, c$_{5}$).}
\label{Kitten_rational}
\end{figure}

In our model we study the emergence of these transitory `kitten' states using the $W$-distribution and its smoothed 
analog the $Q$-function. In the strong coupling limit $\lambda \lesssim 0.05\, \omega$, and  at specific times 
given by the rational submultiples of $T_{\mathrm{long}}: \{T_{p, q} = (p/q)\,T_{\mathrm{long}}|(p,q) = 1,
p \le q\}$  density matrices comprising of a finite number of macroscopic 
coherent states with  uniform angular 
separation on the phase space are observed (Fig. \ref{Kitten_rational}). Starting with the initial hybrid Bell state
 (\ref{t0quasiBell}) of the composite system, the evolution of  $S_{Q}$ 
 in the long range quasi periodic regime 
 shows (Fig. \ref{Wehrl_Time} (b)) the existence of the local minima at rational submultiples of $T_{\mathrm{long}}$.
The presence of \textit{numerous} time scales due to the qubit-oscillator interaction in the present model, however, 
introduces another novel interference related feature. 
In particular, the interaction-dependent \textit{linear} mode with frequency $O(x \widetilde{\Delta})$ causes an 
  energy transfer, in a \textit{short} 
  time scale, between the qubit and  the oscillator degrees of freedom. 
  In the vicinity of the said times 
  $T_{p, q}$ the  oscillations of the  period 
  $T_{\mathrm{long}}$ produce a locally minimum occupation on the phase space,
 whereas the short time period fluctuations  engineer the spread of the occupation by splitting of the Gaussian peaks. This manifests as   
an ordered bifurcation (\textit{\`{a} la} Figs. \ref{Kitten_rational}(a$_{2}$, a$_{4}$) and (a$_{3}$, a$_{5}$), say) 
of the $q$ quasi-probability peaks to $2 q$   peaks representing mixed state oscillator density matrices, while evolving 
from the local minima to the maxima of the \textit{short} time period oscillations of the Wehrl entropy.  
 These \textit{local} variations of $S_{Q}$ in the neighborhood of $(p/q)\,T_{\mathrm{long}}$  are given in 
 Figs. \ref{Kitten_rational} (a$_{1}$)-(c$_{1}$), where we fix $p =1; q=1, 2, 3$, respectively. 
  Moreover, the short range time period ($T_{\mathrm{short}}$) associated with the splitting and subsequent rejoining of the peaks 
 at a particular rational 
 fraction $(p/q)\,T_{\mathrm{long}}$ scales inversely with $q$. For instance, from the Figs. \ref{Kitten_rational}
 (a$_{1}$)-(a$_{3}$), respectively, we observe that $T_{\mathrm{short}} (T_{\mathrm{long}}) \approx 6750, 
 T_{\mathrm{short}} (T_{\mathrm{long}}/2) \approx 3375, T_{\mathrm{short}} (T_{\mathrm{long}}/3) \approx 2250$. This suggests 
 the scaling relation $T_{\mathrm{short}} (T_{\mathrm{long}}/q) \approx 
 (1/q) \,T_{\mathrm{short}} (T_{\mathrm{long}})$. In other words, due to the complex nature of the qubit-oscillator 
 interaction in the strong coupling regime $\lambda \sim 0.05\, \omega$ the exchange of energy is realized between multiple 
interaction-dependent modes, and effectively the high frequency quantum oscillation 
 $O(x \widetilde{\Delta})$ is \textit{frequency modulated} by the low frequency component 
 $O(x^{2} \widetilde{\Delta})$. It is worth mentioning that the oscillator at the local minima 
  of the Wehrl entropy (Figs. \ref{Kitten_rational}(a$_{2}$, a$_{4}$),(b$_{2}$, b$_{4}$), (c$_{2}$, c$_{4}$)) is
 close to pure states. Their respective von Neumann entropy $S$ read $0.26616, 0.17983, 0.53918$, whereas the corresponding maxima
(Figs. \ref{Kitten_rational}(a$_{3}$, a$_{5}$),(b$_{3}$, b$_{5}$), (c$_{3}$, c$_{5}$)) describe almost maximally mixed states. 

\par

Lastly, we note  the geometry of the domain on the phase space that supports the $W$-distribution 
(Figs. \ref{Kitten_rational} (a$_{2}$)-(c$_{2}$), (a$_{3}$)-(c$_{3}$)). The interference pattern realized  between two Gaussian 
peaks  occurs in the intermediate phase space  giving rise to oscillations in a direction perpendicular to the line joining 
the  peaks. Alternate lines with the positive and the negative values of the $W$-distribution appear with 
relative phase differences of $\pi$. As the number of  peaks increase, the interference pattern becomes more complex
while being restricted within a regular polygon with peaks lying at its corners. The Gaussian peaks 
of the $Q$-functions  (Figs. \ref{Kitten_rational} 
(a$_{4}$)-(c$_{4}$), (a$_{5}$)-(c$_{5}$)) appear as smoothed versions of the $W$-distributions. 

\subsection{Wigner entropy and negativity}
\label{Wigner_entropy}

\begin{figure}[H]
\begin{center}
  \subfloat[]{\includegraphics[width=5cm,height=4cm]
 {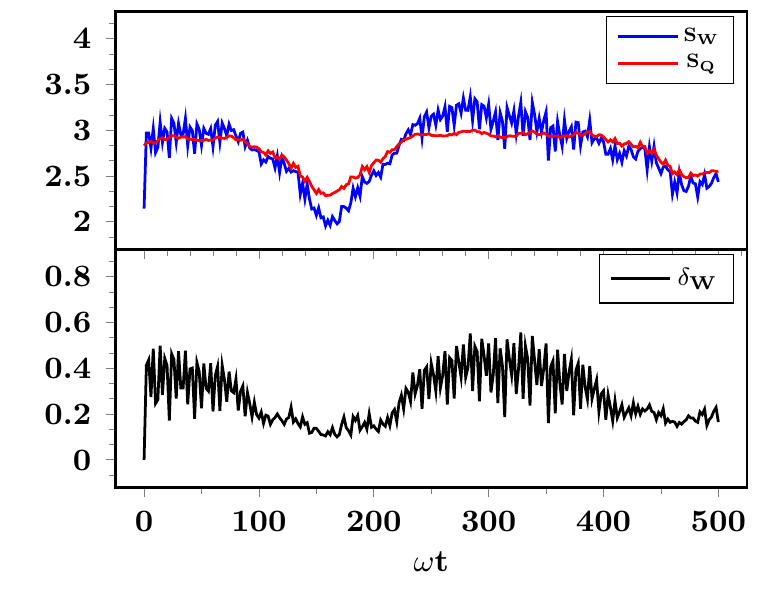}} \quad
\subfloat[]{\includegraphics[width=5cm,height=4cm]
 {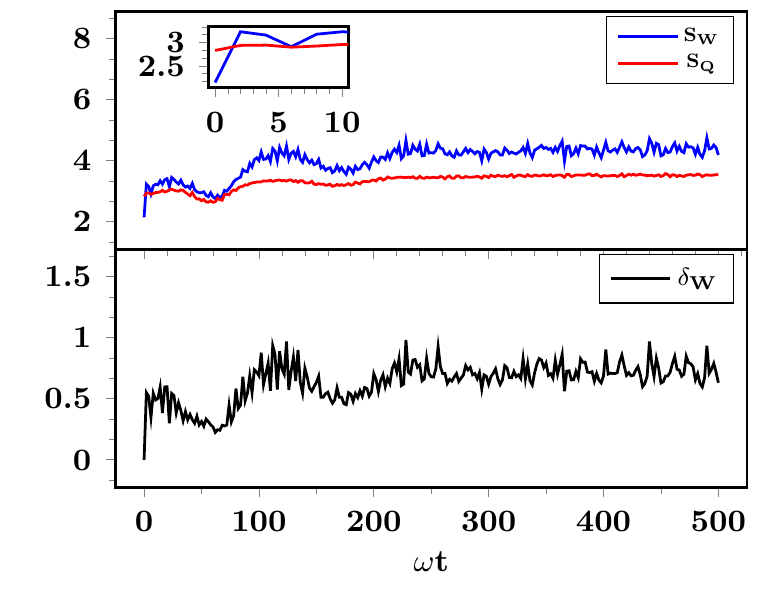}}
 \subfloat[]{\includegraphics[width=4cm,height=4cm]
 {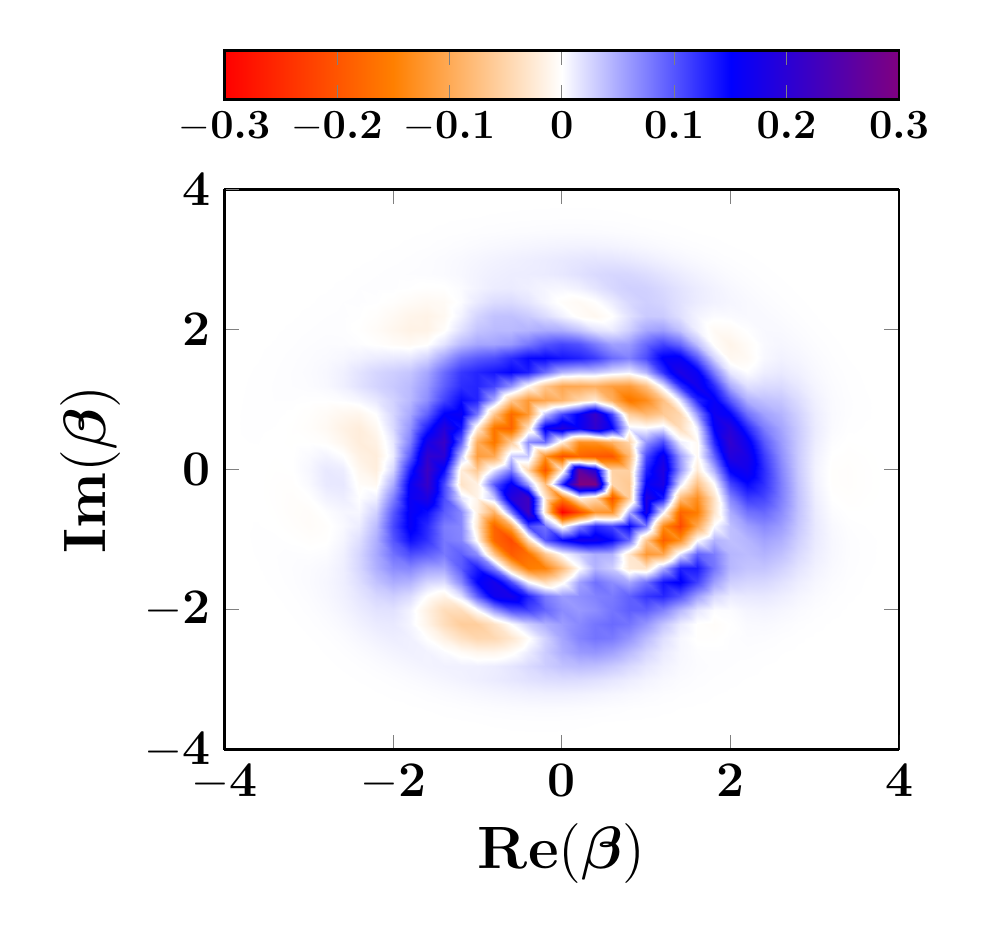}}
\caption{ The time evolution of the Wigner entropy $S_{W}$ (blue), Wehrl entropy $S_{Q}$ (red), and the negativity
$\delta_{W}$ (black) obtained for the parameters $\Delta=0.5\, \omega, \alpha =2$, where the  coupling 
strengths read: (a) $\lambda = 0.1\, \omega$, (b) $\lambda = 0.2\, \omega$. (c) The $W$-distribution in the 
stochastically stabilized domain $(\lambda=0.2\, \omega)$ is given at the scaled time $\omega t=228$, when 
the observed negativity remains prominent: $\delta_{W}=0.9813$.}
\label{Wigner_fig}
\end{center}
\end{figure}

 It is also of interest to study the quantum entropy based on the modulus of the Wigner distribution $|W(\beta, \beta^{*})|$ 
 [\cite{SKD2012}] which is a nonnegative quantity:
\beq
S_{W}=-\int |W(\beta, \beta^{*})| \,\log |W(\beta, \beta^{*})|\; d^2 \beta.
\label{S_W}
\eeq
As the  $W$-distribution contains more information on the phase space structure of a quantum state than its smoothed 
analog the  $Q$-function, a comparative study of the Wigner entropy $S_{W}$ (\ref{S_W}) and the Wehrl entropy $S_{Q}$ 
(\ref{WehrlDef}) is expected to throw a light on the 
nonclassicality of the state. It is evident from the Figs. \ref{Wigner_fig} (a), (b) that the time evolution of the Wigner entropy (\ref{S_W}) and the negativity parameter (\ref{negativity}) have close kinship with each other. This was observed for 
certain oscillator density functions in [\cite{SKD2012}].
 In this sense the Wigner entropy  $S_W$ reveals the extent of nonclassicality of a 
quantum density matrix. We distinguish between two possible scenarios depending upon the qubit-oscillator coupling strength. 
{\sf (i)} In the strong coupling regime $(\lambda/\omega \lesssim 0.1)$ we observe (Fig. \ref{Wigner_fig} (a))  a 
periodic structure that  may be identified with the revival and collapse of the 
qubit density matrix elements reflecting the energy exchanges between the qubit and the oscillator degrees of freedom. 
The collapse, say, of the qubit density matrix elements coincides with a wider spread of the phase space
 distributions of the oscillator. The order of the time period of the revival and collapse of the qubit matrix elements, obtained via retaining up to the linear terms in the Laguerre polynomials, is $O(2 \pi / x \widetilde{\Delta})$.  Therefore 
the  variables such as the  Wigner entropy $S_{W}$, the negativity $\delta_{W}$,
 and the Wehrl entropy $S_Q$ display similar periodic patterns in the said time scale. For a dominant value of  
 $\delta_{W}$ the quantum interference effects are overwhelming, and we, expectedly, find $S_{W} > S_{Q}$, as an increased negativity 
 necessitates an increment in the magnitude $|W(\beta, \beta^{*})|$ for maintaining the normalization property 
 (\ref{W_normalization}). This, in turn, leads to increased value of the entropy $S_{W}$. On the other  hand, for a low negativity domain $\delta_{W} \ll 1$ the inequality is 
 reversed: $S_{W} < S_{Q}$. The underlying reason is that the  $Q$-function is obtained from the $W$-distribution 
(\ref{Q_int_W}) after suitable smearing with a positive definite Gaussian kernel on the phase space, and therefore it 
incorporates less information on the quantum state than the latter [\cite{MF2000}]. {\sf (ii)} In the ultra-strong coupling regime  
$(\lambda/\omega \gg 0.1)$ all modes for the qubit-oscillator interaction with incommensurate frequencies are excited 
and a fully randomized interference pattern  evolves very quickly. Smearing the clear periodic  structures observed earlier 
 these large number of interaction-generated modes lead to quasi stationary values of the phase 
space observables (Fig. \ref{Wigner_fig} (b)). However, despite the statistical stabilization of the occupation on the phase space 
the nonclassicality of the state \textit{remains prominent} due to the interferences occurring between multiple modes.  These 
interferences necessarily develop (Fig. \ref{Wigner_fig} (c)) significant domains on the phase space with  negative values of the $W$-distribution leading to dominant values of the negativity parameter $\delta_{W}$. The average value of  $\delta_{W}$ increases with that of the coupling strength as more interfering modes come into existence. As mentioned before, this results in an   increment  of the entropy $S_{W}$. In the quasi stationary state the 
 negativity $\delta_{W}$ is stochastically preserved. The stochastic stabilization of $\delta_{W}$ occurs after a suitable 
 decoherence time. Therefore in Fig. \ref{Wigner_fig} (b) we observe that except for a brief initial period the Wigner entropy $S_W$ is 
 consistently more than the Wehrl entropy $S_{Q}$, even though the nonnegative $Q$-function may be viewed 
 as a smeared form of the $W$-distribution. Our results (Figs. \ref{Wigner_fig} (a), (b)) suggest that when the negativity $\delta_{W}$
 assumes more than a threshold value  $\delta_{W} \gtrsim 0.4$ the quantum fluctuations ensure the entropy relation: $S_{W} > S_{Q}$.
 
 \par
 
 Another feature revealed in Figs. \ref{Wigner_fig} (a) and (b) is that the relative fluctuations of the Wigner entropy 
 $S_{W}$ overwhelms that of the Wehrl entropy $S_Q: |\delta S_Q/S_Q| \ll |\delta S_W/S_W|$. It signifies that the 
 quantum interferences in the evolution of the  $W$-distribution induce rapid reversal of its sign, whereas the 
fluctuations of different modes are significantly attenuated in the smoothing introduced towards obtaining the  $Q$-function.
 \subsection{Evolution to squeezed states}
 \label{evolution_squeezing}
 
\begin{figure}
\captionsetup[subfigure]{labelformat=empty}
\subfloat[(a)]{\includegraphics[width=4cm,height=3cm]{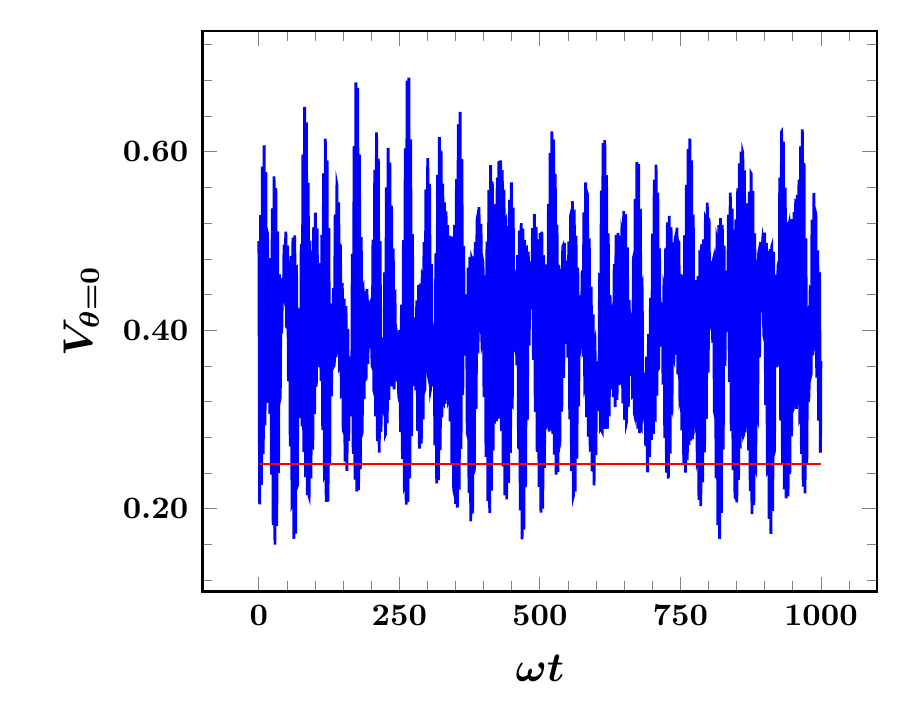}}
\captionsetup[subfigure]{labelformat=empty}
\hspace{-0.3cm}
\subfloat[(b)]{\includegraphics[width=3cm,height=3cm]{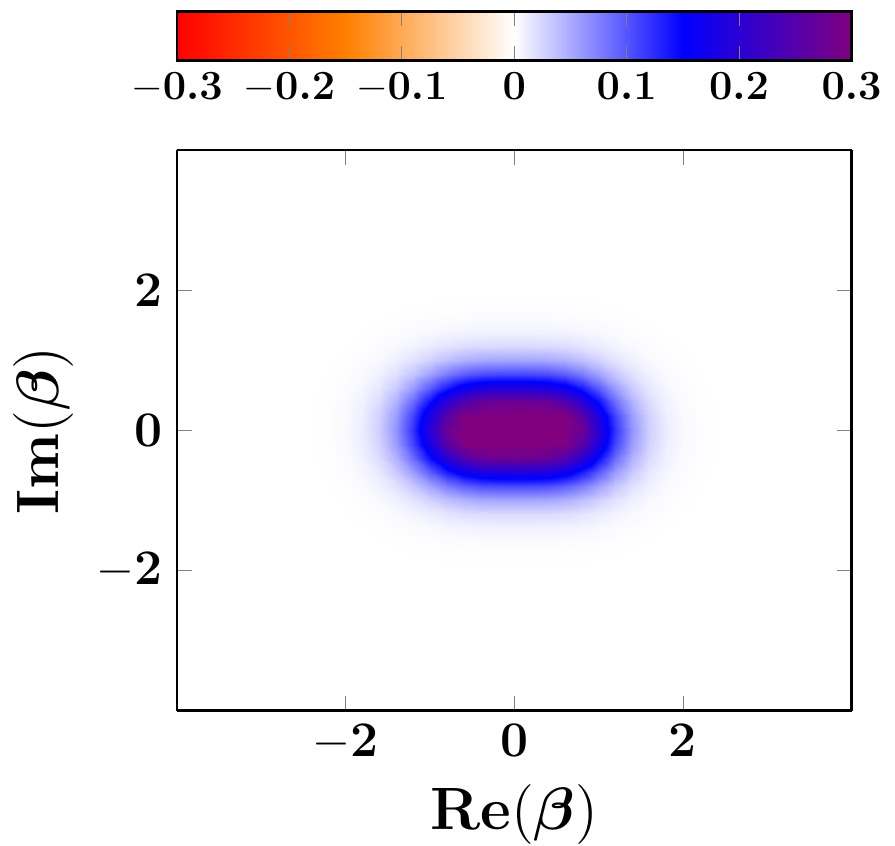}} 
\captionsetup[subfigure]{labelformat=empty}
\hspace{-0.3cm}
\subfloat[(c)]{\includegraphics[width=3cm,height=3cm]{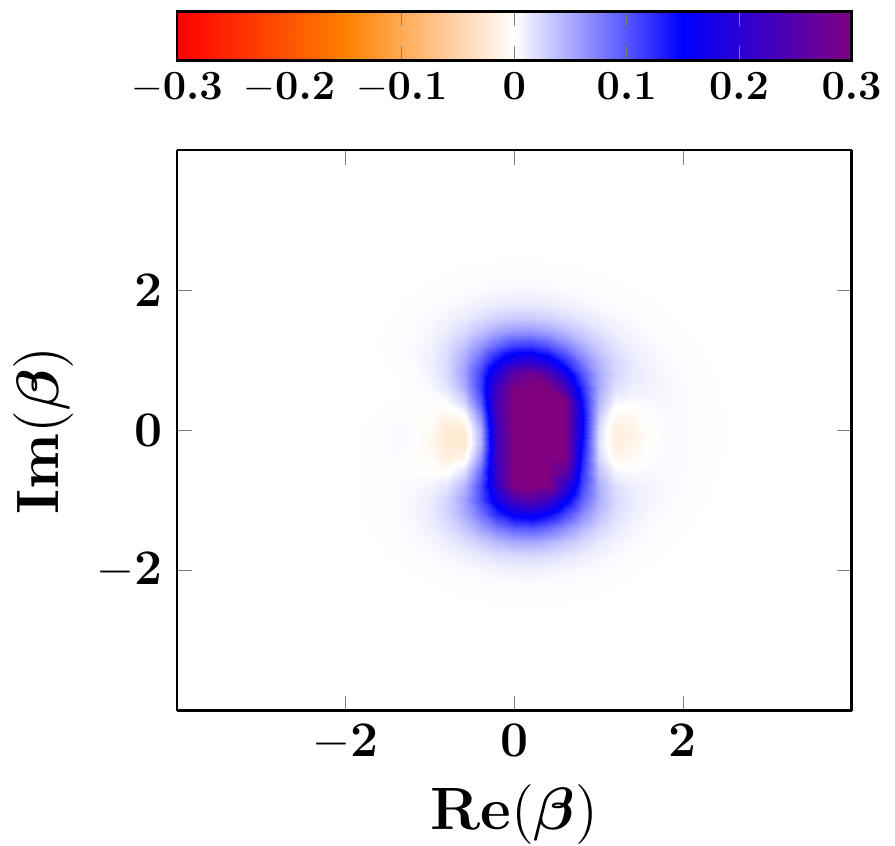}} 
\captionsetup[subfigure]{labelformat=empty}
\hspace{-0.3cm}
\subfloat[(d)]{\includegraphics[scale=0.4]{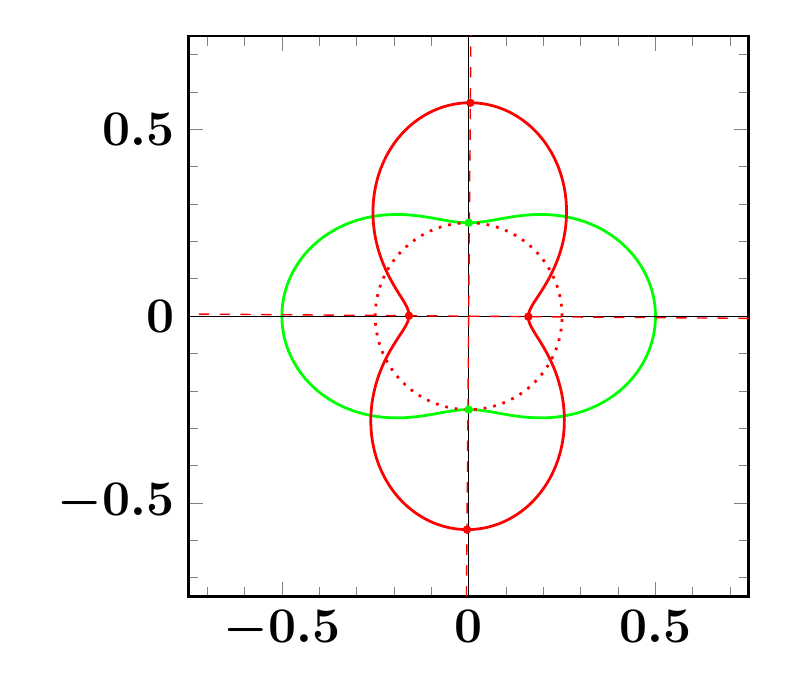}} 
\captionsetup[subfigure]{labelformat=empty}
\hspace{-0.3cm}
\subfloat[(e)]{\includegraphics[width=4cm,height=3cm]{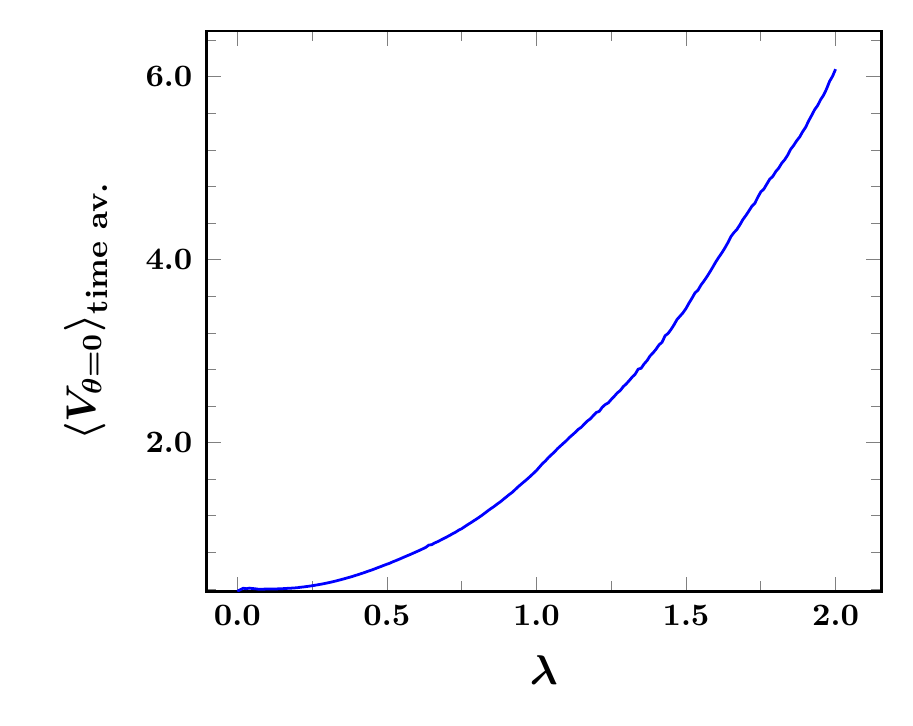}} 
\caption{(a) The time evolution of the quadrature variance $V_{\theta = 0}$ for the parametric choice 
$\Delta= \omega, \lambda=0.1\,\omega, \alpha =0.5$. The horizontal red line represents the limit $V_{\theta}=0.25$.
(b), (c) The  plots of the $W$-distribution at the scaled times $\omega t=0$ and $\omega t=28.80$, respectively. (d) The red dotted circle 
indicates classical limit of the variance $V_{\theta}=0.25$. The polar plots for the variance $V_{\theta}$ at $\omega t=0 \,(\mbox{green})$ and 
$\omega t=28.80 \,(\mbox{red})$ are denoted. The least value of the variance at $\omega t=28.80$ equals $0.15967$ and is observed at 
$\theta=179.55^{\circ}$. (e) The time-averaged  variance $\langle V_{\theta = 0}\rangle_{\mathrm{time\, av.}}$ is plotted w.r.t. $\lambda$
for $\Delta= \omega, \alpha = 0.5$.}
 \label{squeezing}
\end{figure}
A feature of nonclassicality such as squeezing is observed during the time evolution of the state due to the quantum interferences 
between various modes. At relatively small value of the phase space separation $(\alpha \lesssim 1)$ and for the strong coupling regime 
$(\lambda \sim 0.1\, \omega)$, the squeezing is noticed 
  both for the large detuning (Fig. \ref{squeezing}) and the resonant frequencies. The  plot of the $W$-distribution 
(Fig. \ref{squeezing} (c)) makes the quadrature squeezing evident. The signature of the squeezing is observed  
(Fig. \ref{squeezing} (d)) when the variance $V_{\theta}$ of the quadrature variable, say  at $\theta=0$, is rendered less than its
classical value $1/4$. It follows from the polar plot (Fig. \ref{squeezing} (d)) that at the scaled time $\omega t=28.80$
 the quadrature variance $V_{\theta}$ reaches a minimum value $0.15967$ at an angle $\theta=179.55^{\circ}$. The  polar angle  at which 
 the minimum quadrature variance is realized varies with time and depends on the dynamical state of interference of the quantum modes. 
 A qualitative understanding of the squeezing of the state may be described
 as follows. Passing to the interaction picture for the Hamiltonian (\ref{qOH}), an effective Hamiltonian may be obtained \textit{\`{a} la} [\cite{JJ2007}] in an order by order perturbation theory. This effective Hamiltonian contains two photon terms 
$(a^{2}, a^{\dagger\,2})$ at the order $O(x)$ of the coupling strength. These two photon terms give rise to the squeezing of the state. 
An increase in the coupling $\lambda$ first enhances the squeezing as it augments the strength of the two photon terms in the effective 
Hamiltonian. Multiple photon terms, however,  
soon appear in the effective Hamiltonian [\cite{JJ2007}] with increased value of the coupling strength. The resulting randomness 
of the phase relationships of the higher order terms eliminates the squeezing property in the ultra-strong coupling limit. It is obvious in Fig. \ref{squeezing} (a) that the quadrature squeezing, even though it may be present during part of the dynamical evolution of the 
oscillator state, is not realized throughout the oscillatory cycle. The quadrature variance remains above the threshold value:
$V_{\theta = 0} > 1/4$ during the major part of the evolution of the state. The instances of squeezing decreases with increasing 
$\lambda$. To illustrate the feature we plot (Fig. \ref{squeezing} (e)) the time-averaged quadrature variance 
$\langle V_{\theta = 0}\rangle_{\mathrm{time\, av.}}$ 
w.r.t. the coupling strength $\lambda$. We observe that, for the parametric regime studied here $\omega \gtrsim \Delta$,
the time-averaged quadrature variance has no squeezing property: 
$\langle V_{\theta = 0}\rangle_{\mathrm{time\, av.}} > 1/4$, and it smoothly increases with the rising $\lambda$.
The effective Hamiltonian approach [\cite{JJ2007}], however, suggests that the two photon terms $(a^{2}, a^{\dagger\,2})$ underlying the squeezing of the state survive 
the smearing  of the fast oscillatory terms only in the limit of the large qubit frequency: $\omega \ll \Delta$. We will return
to this topic somewhere else.

\par

It is interesting to note that the  squeezed state described in Fig. \ref{squeezing} represents almost pure state of the oscillator 
at the given time as its von Neumann entropy $S\lvert_{\omega t=28.80}\,=\,0.15690$ is much less than its maximal value. The
almost pure state of the oscillator is reciprocated by the corresponding nearly pure state  of the qubit. For the sake of completeness we
include the qubit density matix (\ref{Qubit_DenMa}) at the given time:
\beq
\begin{pmatrix}
 0.90760 & -0.10668 + i \; 0.19300 \\
 -0.10668 - i \; 0.19300  & 0.09240 
\end{pmatrix}.
\label{Qubit_Pure}
\eeq
Its eigenvalues and corresponding  eigenvectors read: 
$\lambda_{1}=0.96342$, $\lambda_{2}=0.03658, \; \ket{\lambda_{1}}=(0.46897- 
i \; 0.84843)\ket{1}-0.24543\ket{-1}$ and $\ket{\lambda_{2}}=(-0.11873+ i\; 0.21480)\ket{1} 
 - 0.96941 \ket{-1}$.
The magnitude of the largest eigenvalue of the qubit density matix 
(\ref{Qubit_DenMa}) may be regarded as measure of the purity of the state. For smaller values of the coherent state amplitude 
  $\alpha \lesssim 0.1$ the purity of the state may exceed $99\%$.  Appearance of a squeezed state requires emergence of appropriate 
 phase relations between various quantum modes that is facilitated by the proximity to a pure state. A statistical mixture of a large number of pure states is likely 
 to destroy the phase relationships  and, consequently, the squeezing property is eliminated.
  % % % % % % % % % % % % % % % % % % % % % % % % % % % %
 \subsection{Mandel parameter and nonclassicality}
 \label{Mandel}
 \begin{figure}[H]
 \captionsetup[subfigure]{labelformat=empty}
 \subfloat[(a)]{\includegraphics[width=4cm,height=2.5cm]{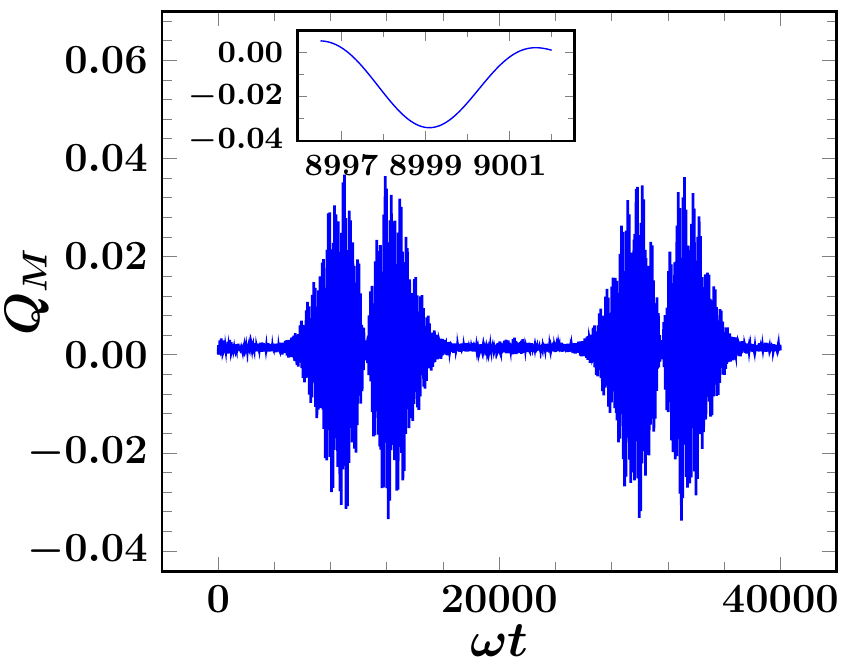}} 
 \captionsetup[subfigure]{labelformat=empty}
 \subfloat[(b)]{\includegraphics[width=4cm,height=2.5cm]{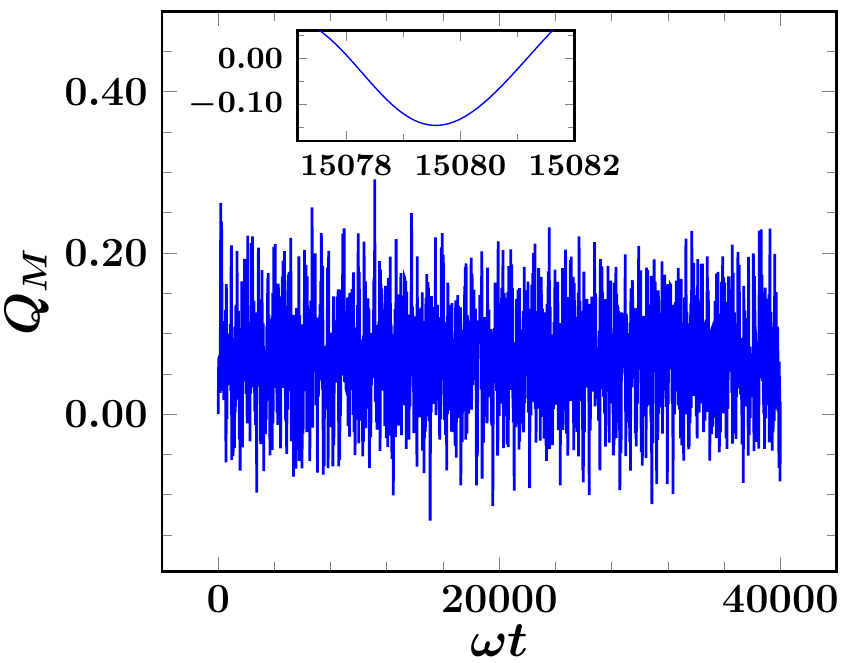}} 
 \captionsetup[subfigure]{labelformat=empty}
 \subfloat[(c)]{\includegraphics[width=4cm,height=2.5cm]{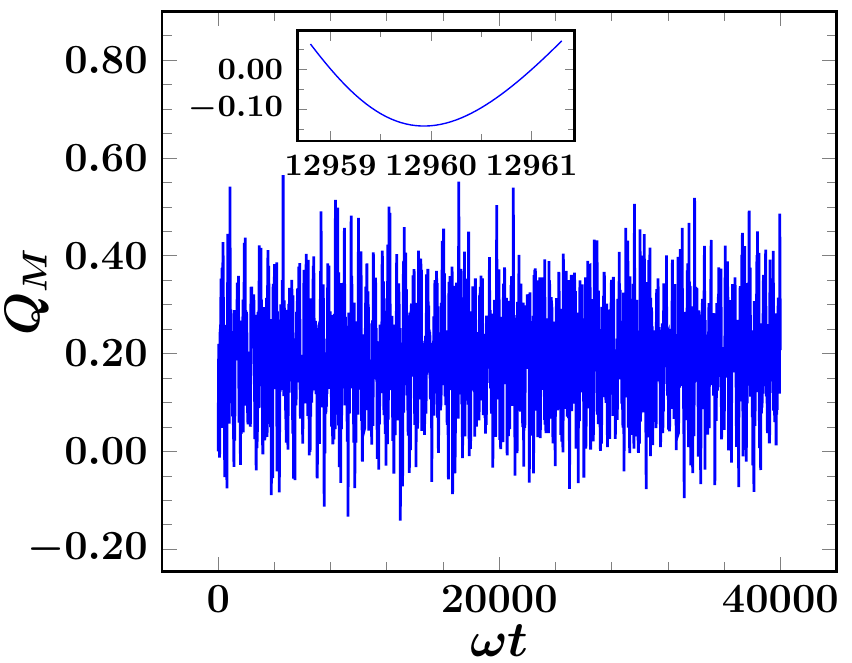}} 
 \captionsetup[subfigure]{labelformat=empty}
  \subfloat[(d)]{\includegraphics[width=4cm,height=2.5cm]{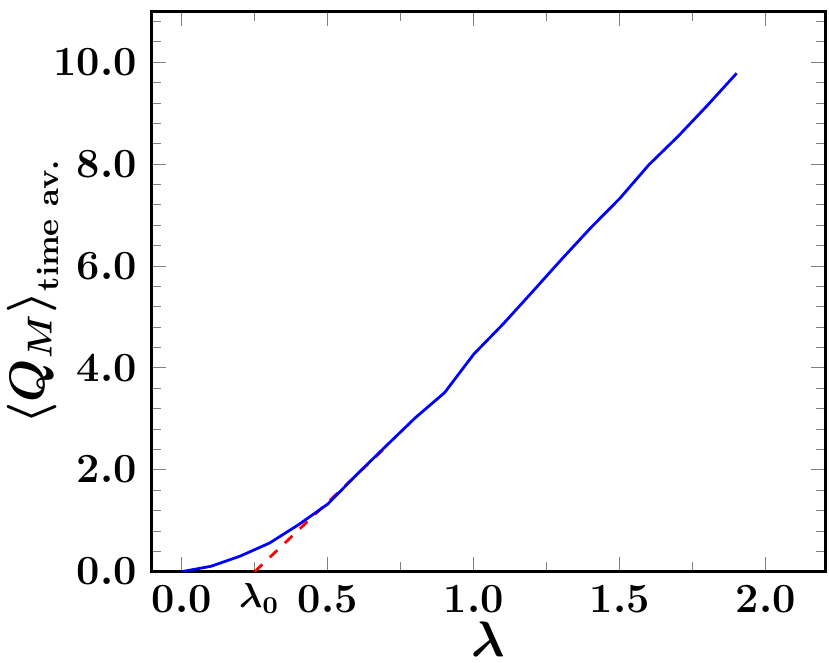}}
 \caption{The evolution of the Mandel parameter $Q_{M}$ for the choice  $\Delta=0.5\, \omega,  \alpha =2$, where 
 the coupling strengths are given by (a) $\lambda=0.01\, \omega$, (b) $\lambda=0.08 \,\omega$, (c) $\lambda=0.15\, \omega.$ 
 (d) The time-averaged value of the Mandel parameter $\langle Q_{M} \rangle_{\mathrm{time \,av.}}$ plotted w.r.t. the coupling strength 
  $\lambda$ for  $\Delta=0.5\, \omega, \alpha =2.$ }
 \label{Mandel_Evolution} 
 \end{figure}
 In experiments allowing a direct detection of photons the Mandel parameter [\cite{M1979}] 
\beq
Q_{M} = \langle (\Delta \hat{n})^{2}  \rangle / \langle \hat{n} \rangle-1
\label{Q_M}
\eeq
is a convenient tool for studying the classical-quantum boundary. For the coherent state the number operator follows the Poissonian statistics  with its signature property $Q_{M} = 0$ that is regarded as the threshold for the classical characteristics. When the quantum correlations in the system suppress the fluctuations in the photon number, it assumes negative values: $Q_{M} < 0$ and  captures the 
sub-Poissonian behavior of the photon statistics that acts as a measure of nonclassicality. The mean value (\ref{avg_photon_number}) and the variance (\ref{variance_num}) of the photon number operator obtained in our model now provide a direct construction of the Mandel parameter $Q_{M}$.  For the coupling strength $\lambda \lesssim 0.1\, \omega$  transient
sub-Poissonian photon statistics  is manifest (Fig. \ref{Mandel_Evolution}) during parts of the evolution of the system. 
The time evolution of  the Mandel parameter $Q_{M}$ in Fig. \ref{Mandel_Evolution} (a), however, passes alternately between the classical 
and the nonclassical regimes. The said sub-Poissonian behavior within short sample intervals is offset by the super-Poissonian photon statistics  $Q_{M} > 0$ during most of the time evolution of the oscillator so that the averaged value over a time scale 
$t_{\mathrm{avg}} \gtrsim 2 \pi/(x \widetilde {\Delta})$ characterising the qubit-oscillator energy fluctuations  always remains nonnegative. The classicality of the photon statistics, therefore, emerge via the time-averaging procedure even though at a shorter time scale the underlying quantum nature of the photon emission process is evident. At higher coupling strengths 
Fig. \ref{Mandel_Evolution} (b), (c) the randomization of the phases generated by a large number of incommensurate modes with the frequencies $\{O(x^{n} \widetilde{\Delta})| n = 1,2,\ldots\}$ partially erase the quantum correlations, while
 making the negativity of $Q_{M}$ less common. The effective Hamiltonian approach [\cite{JJ2007}], as noted in 
 Subsec. \ref{evolution_squeezing}, produce multiple photon terms that give rise to the negativity of $Q_{M}$. Such terms, however, contribute after the smoothing of the rapidly fluctuating components only in the limit $\Delta \gg \omega$.  The time-averaged properties 
the Mandel parameter $Q_{M}$ is summarized in Fig. \ref{Mandel_Evolution} (d), where, in the coupling regime $\lambda \lesssim 0.1\, \omega$ the time-averaged value $\langle Q_{M} \rangle_{\mathrm {time\; av.}} \sim 0$ is maintained so that the overall 
photon statistics effectively remains to be Poissonian. After a brief transition zone the increasing coupling strength 
($ \lambda \gg 0.1\, \omega$) triggers
a full randomization of the  phase relations among the interfering modes, which now constitute an effective bath leading to progressive emergence of classical properties. The resultant stochasticity introduces a scaling behavior for 
the averaged value of the Mandel parameter: 
$\langle Q_{M} \rangle_{\mathrm {time\; av.}} \propto (\lambda - \lambda_{0})$ with a suitable $\lambda_{0}$ that can be 
determined from the Fig. \ref{Mandel_Evolution} (d).
\section{Conclusion}
Employing the generalized rotating wave approximation we have studied an interacting qubit-oscillator bipartite 
system for both the strong and the ultra-strong coupling domains for the choice of an initial hybrid Bell state. 
The evolution of the reduced density matrices of the qubit and the oscillator are obtained via the partial tracing of the
complementary part of the respective degrees of freedom. On the oscillator phase space its density matrix furnishes the diagonal
 $P$-representation that is highly singular due to the presence of rapidly oscillating derivatives of the $\delta$-function.
 Two successive smoothing performed by the Gaussian kernels on the singular $P$-representation produce first the Wigner $W$-distribution,
 and then the Husimi $Q$-function. The quasi probability $W$-distribution admits negative values due to the quantum interferences.
 Its negativity measure $\delta_{W}$ marks the departure of the state from a positive definite distribution on the phase space. The 
 nonnegative $Q$-function provides the Wehrl entropy $S_{Q}$ that acts as measure of delocalization on the oscillator phase space.
 Even though the $W$-distribution may be thought of encapsulating more informations than the $Q$-function, the Wigner entropy 
 $S_{W}$ defined via the magnitude $|W|$ overwhelms the Wehrl entropy $S_{Q}$ whenever the negativity measure $\delta_{W}$
 is dominant. The presence of multiple time scales induced by the interaction introduces a novel feature in the generation of `kitten' states.
 In the coupling range $\lambda \lesssim 0.05 \,\omega$ the long-term quasi periodicity is observed to follow via the  terms 
 $O(x^{2})$ in the interaction Hamiltonian. In this domain the $Q$-function evolves at rational fractions of $T_{\mathrm{long}}$ to a collection of uniformly separated Gaussian peaks representing the `kitten' states.  A shorter time scale $O((x \widetilde{\Delta})^{-1})$ now causes further bifurcation of the Gaussian peaks coinciding with the collapse of the qubit density matrix elements. In the chaotic 
 $\lambda \sim \omega$ regime all interaction-dependent modes of frequencies $\{O(x^{n}\widetilde{\Delta})| n = 0, 1, \ldots\}$ with a
 randomization of their phases. The decoherence time may be estimated as the transient production time of $S_{Q}$ as it approaches its stochastic stabilization. Using the asymptotic behavior of the associated Laguerre polynomials $L_{n}^{(j)}(x)$ the decoherence time is estimated as proportional to $\sqrt{\lambda}$. Nonclassical features such as squeezing and negativity of the Mandel parameter arise due to appearance of multiple photon terms induced by the interaction. In the parametric regime studied here  $\omega \gtrsim \Delta$, such effects do not survive a suitable coarse graining process that smooths the rapidly oscillatory components in the fluctuations.  
 
\label{summary}
\section*{Acknowledgement}
One of us (VY)  acknowledges the support from DST (India) under the INSPIRE Fellowship scheme.

\end{document}